\documentclass[12pt]{article}
\usepackage[T1]{fontenc}
\usepackage{lmodern}
\usepackage{longtable}
\usepackage[font=scriptsize,labelfont=bf]{caption}

\usepackage{array} 
\usepackage{varwidth} 
\usepackage{dsfont}
\usepackage{graphicx}
\usepackage[framemethod=tikz]{mdframed}
\usepackage{amssymb}
 \usepackage[stable]{footmisc}

\DeclareTextFontCommand{\emph}{\em}
\usepackage{amsthm}

\usepackage{epsfig}
\usepackage{blkarray}
\usepackage{multirow}
\usepackage{epstopdf}
\usepackage{hyperref}
\hypersetup{nolinks=true}
\usepackage{verbatim}

\DeclareGraphicsExtensions{.pdf,.eps,.png,.jpg,.mps}
\usepackage{amsmath}
\usepackage{amsfonts}
\usepackage{color}

\usepackage{cite}
\usepackage{fancyhdr}
\usepackage{float}
\restylefloat{table}

\usepackage[utf8]{inputenc}
\usepackage[english]{babel}

\providecommand{\keywords}[1]
{
  \small	
  \textbf{\textit{Keywords ---}} #1
}

\def\lsim{\mathrel{\rlap{\lower3pt\hbox{\hskip0pt$\sim$}}
     \raise1pt\hbox{$<$}}}         
\def\gsim{\mathrel{\rlap{\lower4pt\hbox{\hskip1pt$\sim$}}
     \raise1pt\hbox{$>$}}}         

\begin{document}

\newcolumntype{M}{>{\begin{varwidth}{3.50cm}}l<{\end{varwidth}}} 

\emergencystretch 3em
\begin{titlepage}
\vspace{-3.0in}
\cfoot{\thepage}
\centerline{\Large \bf Hierarchical PCA and Modeling Asset Correlations}
\medskip

\medskip
\centerline{

Marco Avellaneda\footnote{\, New York University (NYU) - Courant Institute of Mathematical Sciences.
Email: \href{mailto:avellane@cims.nyu.edu}{avellane@cims.nyu.edu}.} and 
Juan Andr\'{e}s Serur\footnote{\, New York University (NYU) - Courant Institute of Mathematical Sciences. Email: \href{juan.serur@nyu.edu}{juan.serur@nyu.edu}.}  
}
\bigskip

\begin{abstract}\noindent 
Modeling cross-sectional correlations between thousands of stocks, across countries and industries, can be challenging.  In this paper, we demonstrate the advantages of using Hierarchical Principal Component Analysis (HPCA) over the classic PCA. We also introduce a statistical clustering algorithm for identifying of homogeneous clusters of stocks, or ``synthetic sectors''. We apply these methods to study cross-sectional correlations in the US, Europe, China, and Emerging Markets.
\end{abstract}


\end{titlepage}


\newpage
\section{Introduction and Literature Review}

{} Correlation matrix modeling, one of the main elements in classical portfolio optimization, represents a great challenge as equity markets and their cross-dynamics have become increasingly complex. Traditional empirical estimators have several limitations. For example, as the size of the universe of stocks increases, more observations are needed to estimate the correlations; otherwise, when the number of stocks $N$ is greater than the number of observations $T$, the matrix is singular. Overcoming this with longer estimation windows is generally not an option, as large observation samples are not available, and even if they were, investors prefer shorter estimation windows as stock returns behavior changes dramatically. Furthermore, even if the matrix is not singular, it is usually ill-conditioned with unstable off-diagonal elements. To overcome these problems, practitioners have focused their efforts on modeling correlations using factor models (hence model correlation matrices rather than empirical correlation matrices). 

{}Quantitative factor analysis has become increasingly important in portfolio management. The well-known Capital Asset Pricing Model (CAPM) was one of the most promising breakthroughs in modern finance, proposed in \cite{Sharpe1964}. In this paper, Sharpe proffered a market equilibrium model, where asset returns were explained by the assets' exposure to the market portfolio (systematic risk) plus a idiosyncratic risk component. Intending to generalize the CAPM and accustom it to real-world conditions, \cite{Ross1976} coined the so-called Arbitrage Pricing Theory (APT). Unlike CAPM, this theory is based on arbitrage factor models and extends the equilibrium single-factor model to multi-factor models. In \cite{Fabozzi2010} the authors provide a comprehensive compendium of quantitative methods for equity strategies, with the main focus on portfolio optimization and factor-based models intended to overcome the main shortcomings of the classical Modern Portfolio Theory. In \cite{Avellaneda2010}, the authors show that statistical arbitrage strategies can be formulated using statistical factors (via PCA) and sector ETFs, providing evidence that both approaches work remarkably well and that after augmenting the signals by the traded volume, factor models based on ETFs further enhance their performance.\footnote{For more literature on factor models and their application to risk and portfolio management, see \cite{Fabozzi2006}, \cite{Fabozzi2007}.} 

{}Developing systematic portfolio strategies is now regarded as the core intellectual activity of ``quant funds'' managing hundreds of billions of dollars in assets. Usually, practitioners choose between two types of factor models. On the one hand, there are models based on explicit factors such as momentum, value, size, quality, among others. In \cite{Fama1992} and \cite{Fama1993} the authors proposed a multi-factor model, extending the CAPM by adding the factors value and size. Empirical evidence has shown that this model provides a better characterization of the cross-section of the stocks' returns. More recently, Fama and French came up with a new model (see \cite{Fama2015}), augmenting its previously three-factor model with the factors profitability and investment, originating the so-called  Fama and French five-factor model. 

{}On the other hand, there are those models based on implicit factors like statistical features extracted from assets' returns using Principal Component Analysis (PCA), maximum likelihood, among others approaches. For example, in \cite{Connor1988} the authors, using asymptotic PCA, identified five factors that work remarkably better than CAPM and are able to capture the time-varying risk premium.\footnote{For more recent developments in this arena, see, for example, \cite{KakushadzeYu2017}, \cite{Meucci2010}, \cite{Torun2011}.} 

{}One of the main advantages of implicit factor model is that they do not make assumptions about the drivers behind price movements. They rely on market data without additional information. However, like any technique in finance, they are not a panacea. These models must be treated carefully to avoid undesirable instabilities that lead to high estimation errors. In addition, there are several challenges. For example, setting up the number of $K$ implicit factors --mainly in the context of PCA-- is a crucial step. In this matter, various techniques have been proposed. One of the most famous approaches is based on Random Matrix Theory (RMT) (see \cite{Cizeau2000} and \cite{Laloux2000}), intended to retain only a few significant eigenvectors and filter out the noisy ones, modeling them as random noise. In \cite{Kakushadze2015}, the author proposed the ``minimization algorithm'', choosing the number $K$ based on the total risk attributed to idiosyncratic risk as $K \to N$, where $N$ is the number of stocks. So the approach aims to minimize the absolute difference between function $g(K)$ and 1.\footnote{Here, the function $g(K) = \sqrt{\min(\hat\xi^2)} + \sqrt{\max(\hat\xi^2)}$, where $\hat\xi$ is the idiosyncratic risk.} Likewise, the effective rank proposed in \cite{Roy2007} is a versatile method based on the Shannon entropy, intended to measure the effective dimension of the matrix as it is generally lower than the number of positive eigenvalues due to the highly correlated components.

{}Throughout this paper, we implement a technique called Hierarchical PCA (HPCA), introduced by \cite{Avellaneda2019} in the context of equity correlation matrices ordered hierarchically by the MSCI Global Industrial Classification Standard (GICS). To model this hierarchical structure, we use the GICS and countries. Furthermore, we present a novel statistical clustering technique that harnesses the power of PCA, which is based solely on returns data and is capable to overcome some drawbacks inherent in static-like clusters such as GICS and/or countries. Empirical analysis shows that the eigenvectors obtained by HPCA works outstandingly well, overcoming some issues of classical PCA. Finally, we provide some trading strategies ideas leveraging the modeled factors. To make results fully reproducible, we repeat the analysis to the US, European, Chinese, and Emerging stock markets.

\section{PCA revisited} \label{FirstHP2}

{}In a universe consisting of $N$ stocks and $T$ observations, we consider the $N\times N$ empirical correlation matrix,
\begin{align}
& C = \frac{1}{T} RR^t \ 
\end{align}
{}where $R$ is the $T\times N$ matrix of standardized returns.\footnote{As usual, standardized return = $\frac{r_t-\overline{r}}{\sigma(r)}$.}

{}PCA calculates the eigenvalues and eigenvectors of the correlation matrix ranked in decreasing order by eigenvalues. Accordingly, the first eigenvector solves the variational problem
\begin{align}
& V^{(1)} = \textit{argmax} ~\{V^t C V: ||V||_2 = 1\} \ \label{Eigen1}
\end{align}
{}where $||.||_2$ represents the Euclidean space.\footnote{The $n$-dimensional Euclidean norm in $\mathbb{R}^n$ is defined as $||X||_2 := \sqrt{X_1^2 + X_2^2 + X_3^2 + \dots + X_m^2}$.} Higher-order eigenvectors satisfy a similar variational problem, restricting the problem to the orthogonal complement to the previous eigenvectors:
\begin{align}
& V^{(k)} = \textit{argmax} ~\{V^t C V: ||V||_2 = 1, V^{(k)t} V^{(r)} = 0, ~1 \leq r < k \} \ \label{Eigen2}
\end{align}

{}The vectors $V^{(k)}$ satisfy $CV^{(k)} = \lambda^{(k)}V^{(k)}$, i.e. they are eigenvectors of $C$,
which can be factored as
\begin{align}
& C = \mathcal{V} \Lambda \mathcal{V}^T. \
\end{align}
{}Here $\mathcal{V}=[V^{(1)},...,V^{(N)}]$ is an orthogonal matrix. Its columns of which are the eigenvectors of $C$, and $\Lambda$ is the diagonal matrix with the eigenvalues ordered from the highest to the lowest.\footnote{Note also that the eigenvectors are defined up to sign. For more details, see \cite{Jollife2002}. }$^,$\footnote{In financial economics, the first-order optimality condition for a portfolio that maximizes the Sharpe Ratio over all competing portfolios investing in the same $N$ stocks can be represented as $r - E(r) = \beta_r (F- E(F)) + \epsilon_r$. Thus, remarkably, the Principal eigevector is connected to the concept of the Market Portfolio in the sense of Modern Portfolio Theory \cite{Avellaneda2010}, and \cite{Boyle2014}.}

We define the {\it $k^{th}$ principal eigenportfolio} as the portfolio with loadings
\begin{equation}
\theta^k_i = V^{(k)}_i/\sigma_i \, \ \ \ i=1,2,...N,
\end{equation}
where $\sigma_i$ represents the standard deviation of the returns of asset $i$. 

\subsection{Eigenvalue Analysis and Spectral Cutoffs}

\begin{figure}[H]
\includegraphics[width=0.90\linewidth]{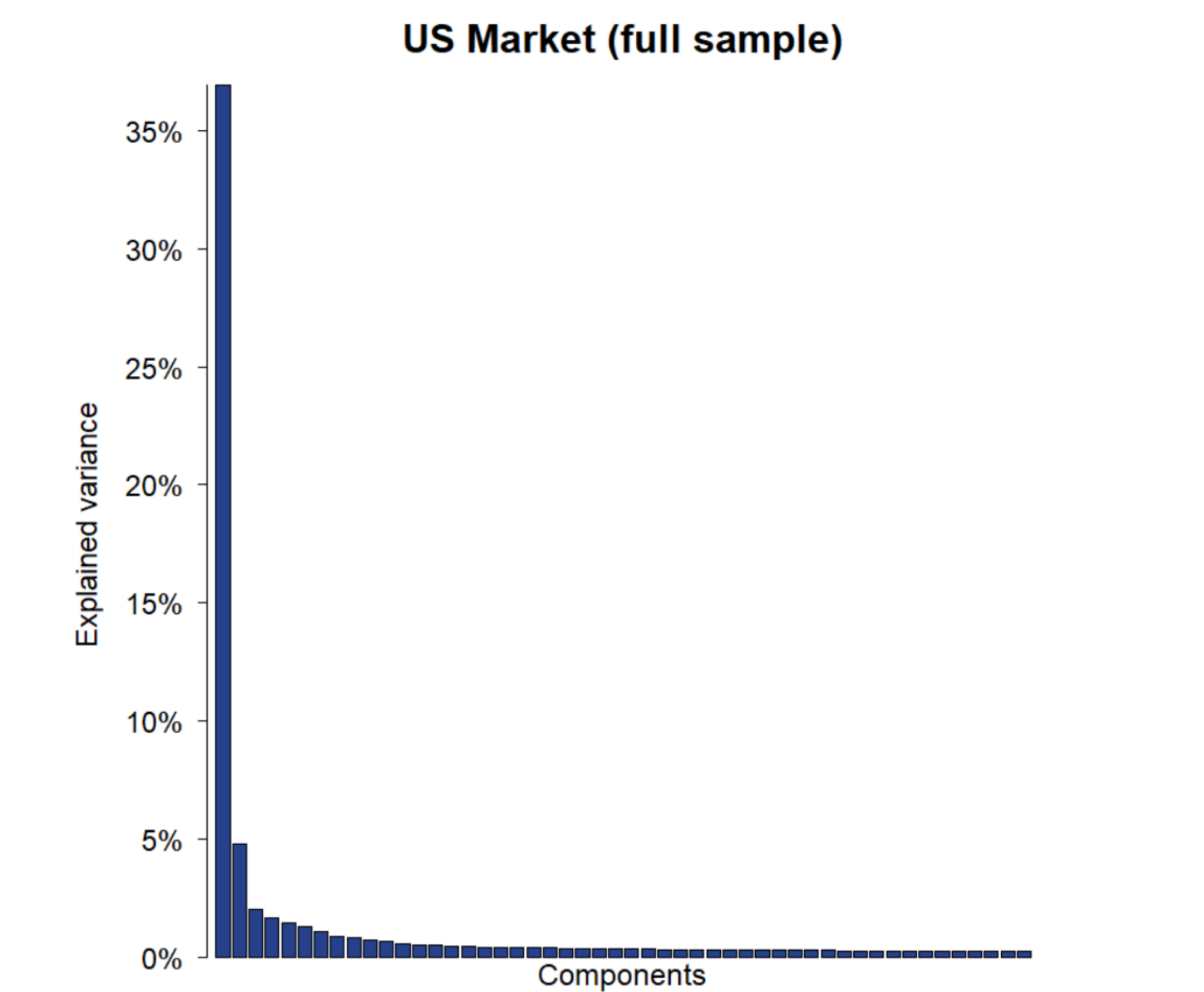}
\caption{Top 50 eigenvalues for the US stocks with the data sample spanning from 2010 to 2019. The variance explained by the first eigenvalue is approximately $\lambda^{(1)}/N$ = 40\%. }
\label{spectraEU}
\end{figure}

{} We see from the above that PCA is a ``greedy algorithm'' in the sense that the eigenvectors (or, more precisely, eigenportfolios) explaining the largest market variability are extracted sequentially. Figure \ref{spectraEU} displays the top 50 eigenvalues of the correlation matrix of US stocks' returns.\footnote{for the constituents of the S\&P 500 Index}.

\begin{figure}[H]
\includegraphics[width=1\linewidth]{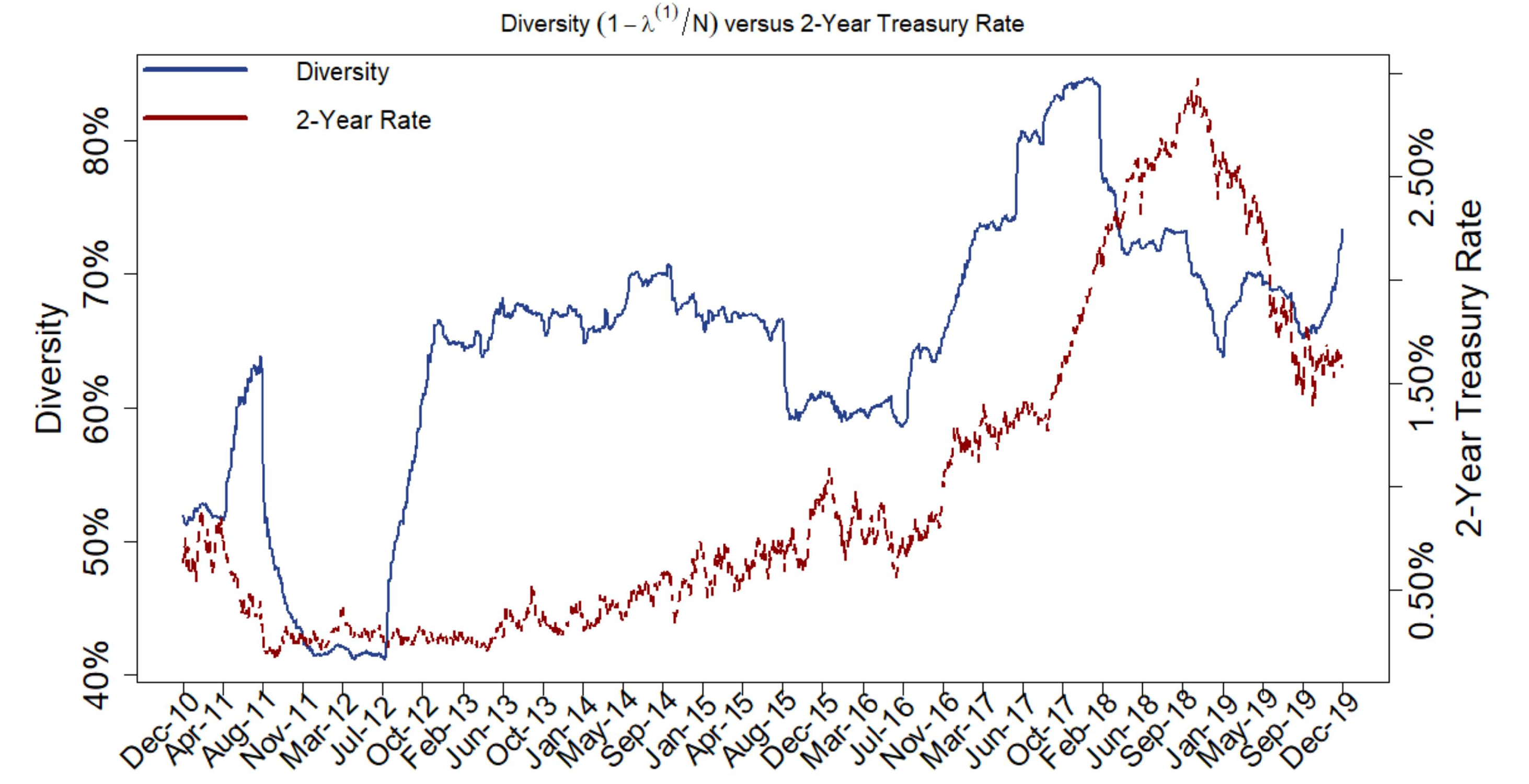}
\caption{Diversity level ($1 - \lambda^{(1)}/N$) is calculated with the first eigenvalue using a 1-year rolling correlation matrix. The diversity level moves in the same direction as the 2-Year Treasury Constant Maturity Rate. The 2-Year CMT Rate was obtained from the FRED repository.}
\label{TimeSerieEigen2}
\end{figure}

Figure \ref{spectraEU} shows that the first eigenvalue accounts for almost 40\% of the variance. However, it is well-known empirically that the amount of variance\footnote{This is actually measured as a percentage of the total trace.} explained by the principal eigenportfolios, or by the top eigenvectors, varies over time \cite{Avellaneda2010}. In financial stress periods, the variance explained by the first eigenportfolio increases sharply, depicting the increase in correlations across different assets and attempting against diversification benefits. This is shown in Figure \ref{TimeSerieEigen2}, which compares the ``diversity level'' with the 2-year Treasury Rate.\\

{} Ill-conditioning of correlation matrices for large groups of equities is a feature of the market: the ratio of the largest to the smallest eigenvalues (condition number) is very large.  If we fixed the number of observation dates $T$, and gradually increase the number of stocks $N$, the more correlated stocks would appear to be.

{} In practice, short observation windows are necessary. Investment managers typically ``refresh'' correlation matrices and use a relatively short window for sampling data, such as 120 or 180 days. For large universes, this leads to degenerate ``empirical'' correlation matrices. 

{} The correlation coefficients of stocks which are not linked, economically or otherwise, are unreliable, or at least suspicious. Practitioners often use the eigenvalues of $C$ to determine what they believe are ``significant factors'', attempting to separate separate ``signal'' from ``noise'' and regularize degenerate or ill-conditioned correlation matrices.  This leads naturally to dimension-reduction techniques, such as using spectral cutoffs that retain only a few ``significant'' eigevectors and model the orthogonal complement as random noise (see \cite{Ledoit2014}), some of which are based on RMT, introduced in Physics by Wigner in 1930 (see, for example, \cite{MarcenkoPastur}, \cite{Laloux2000}).

\subsection{Higher-order eigenvectors and eigenportfolios: the identification problem}

\begin{figure}[H]
\includegraphics[width=1\linewidth]{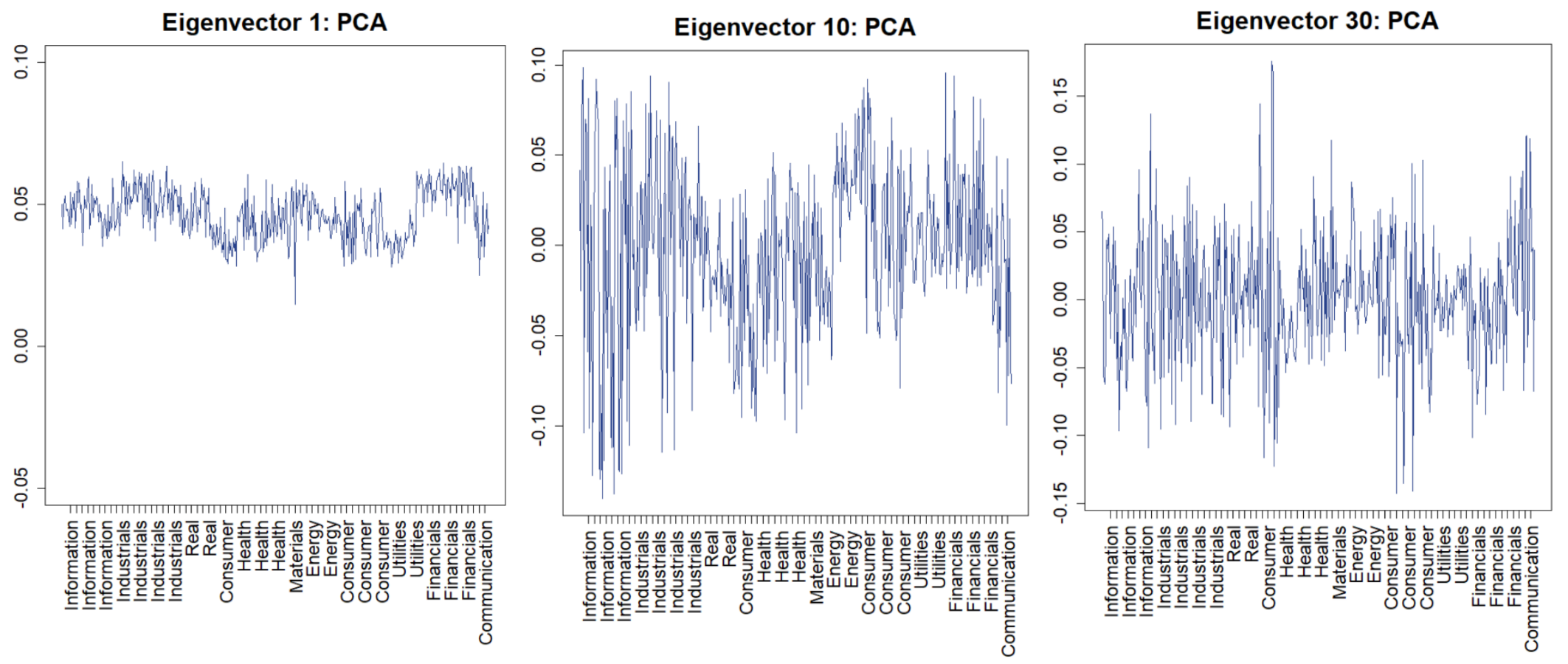}
\caption{PCA eigenvectors for US markets. The first represents the market, but those of higher-order (tenth and thirtieth in this case) suffer the so-called identification problem since it is very difficult to find a meaningful economic intuition.}
\label{interpret}
\end{figure}

{}A particular issue with PCA is that beyond the first eigenportfolio --associated with the market mode-- it is difficult to find an economic or financial explanation in higher-order eigenportfolios which are believed to be ``significant'' after applying a spectral cutoff \cite{Laloux2000}; see Figure \ref{interpret}. This is quite different from the standard situation in fixed-income, where changes in  bond yields or forward rates can more easily be interpreted in terms of higher-order eigenportfolios (curve steepening/flattening, or changes in 2-versus-10 or 10-versus-30 yields \cite{Litterman1990}).

{} In Figure \ref{interpret}, the first eigenvector (and eigenportfolio) is straightforward to  interpret. It has all the positive weights, and is a reasonable proxy for the ``market portfolio'', such as an S\&P 500 index tracker. However, other eigenvectors/eigenportfolios are less straightforward to interpret. They have positive and negative weights across the spectrum, making it difficult to analyze for portfolio management purposes; see nevertheless \cite{GopiStanley} and \cite{Avellaneda2010}.

\section{Hierarchical PCA}

{} As a first step, we partition the stock universe into clusters which are believed to share common features, such as an industry sector or a country, or that stocks are otherwise clustered according to the modeler's beliefs. The exact specification of the clusters will be addressed later. 

{} We assume that the modeler holds strong beliefs on the data but only for stock returns belonging to the same cluster. In contrast, the modeler trusts less the empirical correlations between stocks which belong to different clusters. To fix ideas, assume that there are $b$ clusters.

In addition, assume that the modeler chooses $b$ ``benchmark portfolios'', each of which is associated with a cluster. The benchmark could be, for instance, an ETF tracking a basket of stocks in the cluster. The modeler believes in the correlations between the pairs of the benchmark portfolio returns. We denote the correlations by $\rho^{k,k'};\ k=1,..,b,\ \ k'=1,..,b$.

Let $F^k$ denote the standardized returns
of the benchmark portfolio associated with cluster $k$. The regression coefficient of stock $i$ on the return of the corresponding benchmark portfolio, $F^k$, is denoted by $\beta_i$. We assume that the modeler also believes in the $\beta_i$.

{} Consider the function $\mathds{I}(i)$, which returns the sector of stock $i$: so $\mathds{I}(i) = \mathds{I}(j)$ if and only if asset $i$ and the asset $j$ belongs to the same sector. 

A correlation matrix $\hat{C}$ which 
incorporates the modeler's beliefs in a parsimonious fashion is
given by:
\begin{align}\label{Cases}
 \hat{C}_{i, j} =
  \begin{cases}
    C_{i, j} & \text{if} ~~~\mathds{I}(i) = \mathds{I}(j)\\ 
   \beta_i \beta_j \hat{\rho}^{\mathds{I}(i), \mathds{I}(j)} & \text{otherwise}.
  \end{cases}
\end{align}

{} In fact, the reader can may easily check that $\hat{C}$ represents the correlation matrix of a Gaussian probability measure in $N$-dimensions.\footnote{This probability measure is the  maximum-entropy distribution, in which the modeler's beliefs are viewed as moment constraints \cite{GolanJudgeMiller}.}$^,$\footnote{Justifying the claim that the model is ``parsimonious''. It is readily seen that $\hat{C}$ is indeed non-negative definite with unit diagonal elements \cite{Avellaneda2019}, and thus corresponds to a correlation matrix.}

\subsection{Selecting the benchmark portfolios as the first eigenportfolios for each cluster}

A further simplification comes from making a specific choice: we assume that the benchmark portfolio for a given cluster
is the first eigenportfolio of the cluster. The standardized return of the benchmark portfolio
of sector $k$ can be written in the form 
\begin{align}
& F^{k} = \frac{1}{\sqrt{\lambda^{1,k}}} \sum_{i: \mathds{I}(i)=k} V_i^k X_i\
\end{align}
{}where $X_i$ represents the standardized returns of stock $i$,
$V^k_i$ is the first column in the PCA factorization of the correlation matrix of sector $k$, and $\lambda^{1,k}$ is the first eigenvalue. Notice that the benchmark portfolios are standardized by definition (mean = 0, variance = 1).

From the orthogonality of the eigenportfolios in the same sector, we can derive a simple formula for the regression coefficients:

\begin{equation}
\beta_i = Corr(X_i, F^{\mathds{I}(i)}) = \sqrt{\lambda^{1,\mathds{I}(i)}}\ V^{\mathds{I}(i)}_i.
\label{beta}
\end{equation}

\begin{figure}[H]
\includegraphics[width=1.\linewidth]{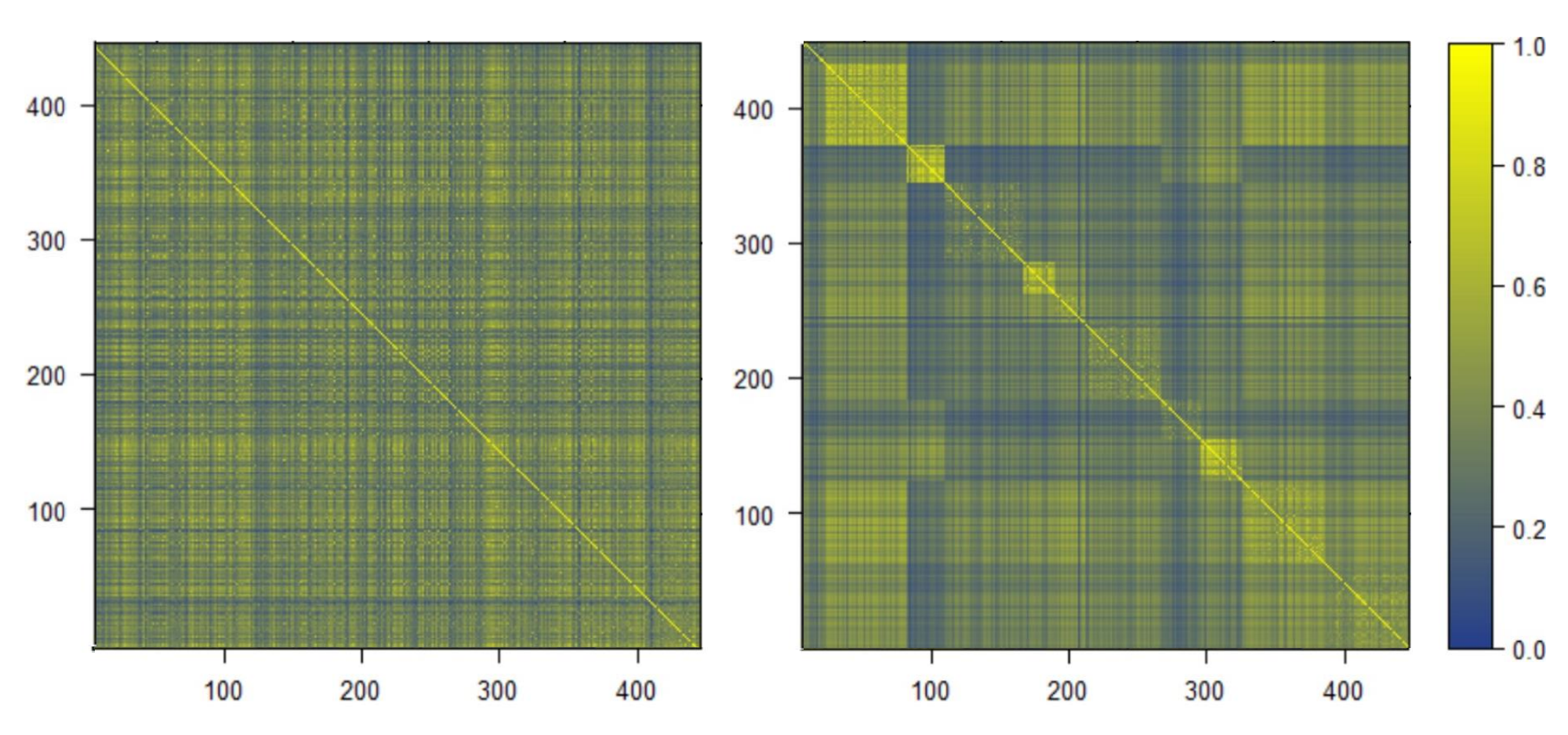}
\caption{Original (left) and modified (right) correlation matrices estimated with the S\&P 500 returns' constituents and GICS clusters, from 2010 to 2019. Dark areas mean low correlation, while lighter areas mean higher correlation.}
\label{hpca}
\end{figure}

{} Figure \ref{hpca} displays the empirical correlation matrix $C$ side-by-side with the model correlation matrix $\hat{C}$ (HPCA matrix). The HPCA matrix presents a more clear, distinct block structure. The blocks associated with inter-sector correlation are lighter, suggesting that correlations between stocks in different clusters are less pronounced than in the empirical data.

\section{Examples}

\subsection{Data and Methodology}

{} We analyze four major global equity markets: the United States, Europe, Emerging Markets and China.\footnote{The data consists of end-of-day prices adjusted for dividends and splits extracted from the Reuters Market Data System. To ensure homogeneity of the asset returns and avoid differences due to the currency of each country, all asset prices were converted to US dollars before calculating the stock returns.} The data covers the period from January 2010 to November 2019. This period includes different macroeconomic events which affected world markets, such as the ``flash crash'' of 2010, the European financial crisis in 2011/2012, the downgrade by Standard and Poor's of the U.S. Treasury, the invasion of Ukraine, the Ebola virus outbreak, the downgrading of the Chinese Yuan in 2015, Brexit in 2016, the U.S. elections of 2016, and the Trade Wars of the end of the decade.\\
\medskip
\begin{table}[H]
\centering
\begin{tabular}{clcccc}
  \hline
 & Sector (GICS) & USA & Europe & Emerging Mkts. & China \\ 
  \hline
& Communication          & 24 & 42  & 59  & 10  \\ 
& Consumer Discretionary & 64 & 66  & 114 & 84 \\ 
& Consumer Staples       & 32 & 43  & 92  & 23  \\ 
& Energy                 & 28 & 22  & 56  & 5  \\ 
& Financials             & 64 & 109 & 277 & 19  \\ 
& Health Care            & 60 & 54  & 54  & 50  \\ 
& Industrials            & 69 & 115 & 89  & 97  \\ 
& Information Technology & 69 & 33  & 121 & 88  \\ 
& Materials              & 28 & 49  & 121 & 81  \\  
& Real Estate            & 31 & 33  & 44  & 22  \\ 
& Utilities              & 28 & 30  & 46  & 19  \\
   \hline
&\textbf{Total} & \textbf{497} & \textbf{596} & \textbf{1049} & \textbf{498}\\
   \hline
   \hline
\end{tabular}
\caption{\label{Data} Numbers of companies considered in the study by GICS sectors and regions.}
\end{table}

{}Table \ref{Data} shows the main 11 GICS sectors and the number of companies in each sector by geography.  To give  perspective on the countries included in the European and Emerging markets groups, Table \ref{DataCountries} in Appendix shows the number of companies belonging to the main countries for these regions.

{}In the following sections, we apply HPCA for the four regions mentioned above. As stated in the introduction, the standard (static) clustering method is based on the GICS. In addition, Emerging and European markets are also clustered by countries. Finally, we propose a statistical clustering technique that is based solely on asset returns and not on exogenous information.

\subsection{US Stocks}

{}The stocks analyzed correspond to the S\&P 500 constituents, clustered according to the GICS metric. The main GICS are Information Technology, Industrials, Financials, Consumer Discretionary, and Health Care, accounting for more than 65\% of the US stocks.

{}The reason to use the GICS is to capture the economic link between each sector, which is appealing if we consider that companies belonging to the same industries share common factors that can be identified and provide insight on their behavior.

\subsubsection{Eigenvalue Analysis of the HPCA: clusters based on GICS} 

\begin{figure}[H]
\includegraphics[width=1.\linewidth]{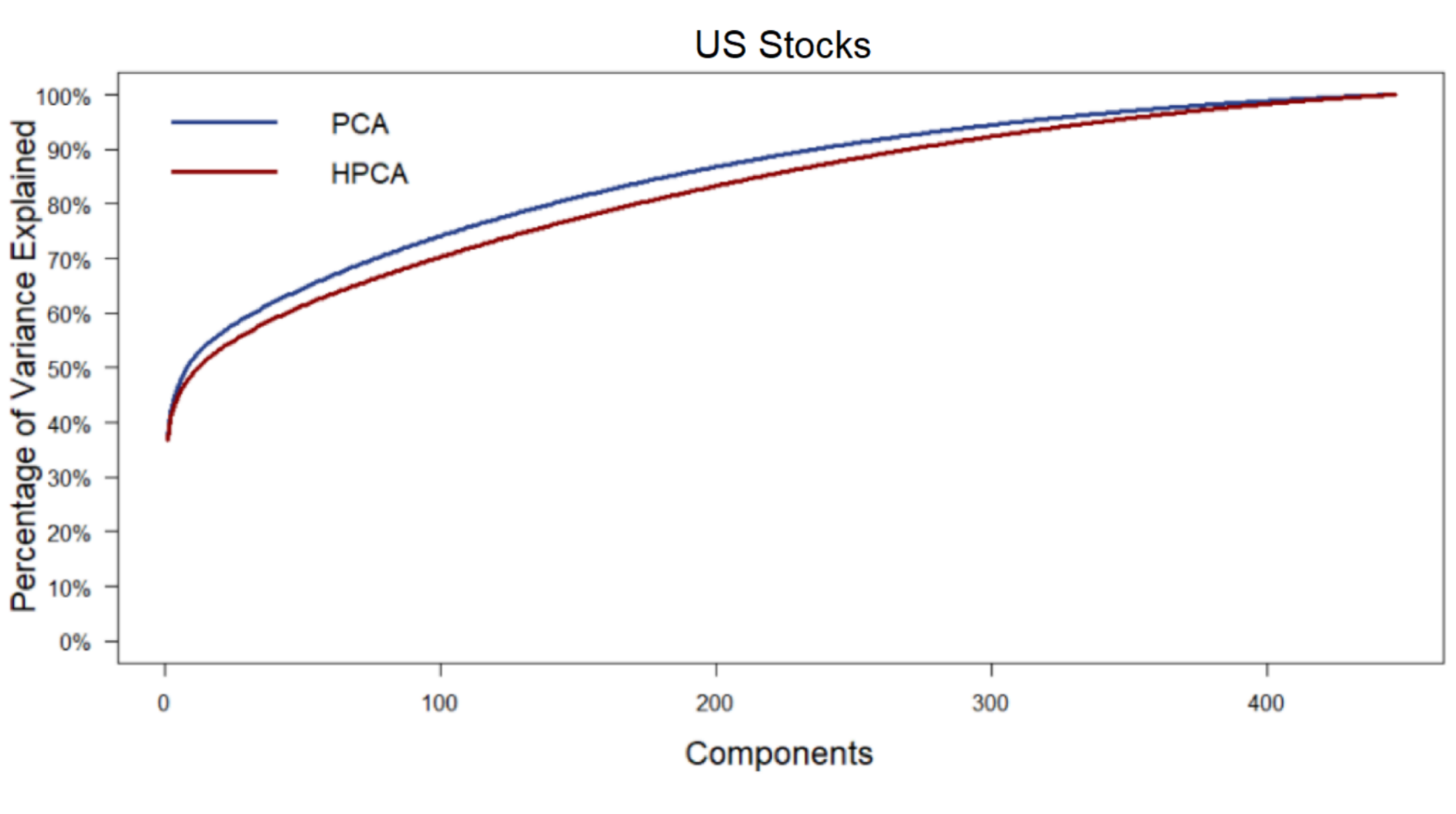}
\caption{\label{USHPCA}Cumulative variance explained by eigenvalues of PCA and HPCA.}
\end{figure}

{}Figure \ref{USHPCA} shows that the curve of cumulative explained variance of PCA rises faster than the curve of HPCA due to the nature of HPCA algorithm. This is considered good since it indicates a lower concentration in a few components. In other words, HPCA is less ``greedy'' than PCA.\\

{}Table \ref{Data3} depicts the same as shown in the previous figure. HPCA has lower eigenvalues than PCA across the spectrum.

\begin{table}[H]						
\centering						
\addtolength{\tabcolsep}{-3pt} 
\begin{tabular}[t]{clccc|ccc|llc}						
   \hline						
   \multicolumn{6}{r}{US Stocks}\\ 
   \hline						
  \hline						
&Eigen	&&PCA	&HPCA	&PCA (\%) &HPCA (\%) &&& Eigenportfolio\\  \hline	
  \hline					
& Eigen 1	&&164.65	&163.55	&37.01\% &36.67\% &&& Multi-sector\\		
& Eigen 2	&&21.24	&18.55	&4.77\%	 &4.16\% &&& Multi-sector\\
& Eigen 3	&&8.89	&7.10	&1.99\%	 &1.59\% &&& Multi-sector\\		
& Eigen 4	&&7.40	&6.18	&1.66\%	 &1.39\% &&& Multi-sector\\		
& Eigen 5	&&6.43	&5.41	&1.44\%	 &1.21\% &&& Multi-sector\\		
& Eigen 6	&&5.61	&4.27	&1.26\%	 &0.96\% &&& Multi-sector\\		
& Eigen 7	&&4.81	&3.92	&1.08\%	 &0.88\% &&& Multi-sector\\		
& Eigen 8	&&3.88	&3.04	&0.87\%	 &0.68\% &&& Multi-sector\\		
& Eigen 9	&&3.51	&2.81	&0.78\%	 &0.63\% &&& Consumer Disc.\\		
& Eigen 10	&&3.13	&2.70	&0.70\%	 &0.61\% &&& Multi-sector\\		
& Eigen 11	&&2.82	&2.56	&0.63\%	 &0.57\% &&& Financials\\		
& Eigen 12	&&2.44	&2.53	&0.55\%	 &0.57\% &&& Inf. Technology \\		
& Eigen 13	&&2.25	&2.19	&0.51\%	 &0.49\% &&& Health Care \\		
& Eigen 14	&&2.13	&2.16	&0.48\%	 &0.48\% &&& Health Care\\		
& Eigen 15	&&2.06	&2.07	&0.46\%	 &0.46\% &&& Industrials\\		
\hline						
\hline						
\end{tabular}						
\caption{\label{Data3} First 15 HPCA and PCA eigenvalues clustered by GICS for the US market. In the Eigenportfolio column, we show if a given eigenportfolio is constituted by multiple or single sectors.}						
\end{table}						

{}Furthermore, we identified two types of eigenportfolios: those localized in multiple sectors and those concentrated in a single sector. Table \ref{Data3} shows that almost all eigenportfolios from the first to the fifteenth are multi-sector.

\subsubsection{Eigenvector Analysis of the HPCA: clusters based on GICS}

\begin{figure}[H]
\includegraphics[width=1\linewidth]{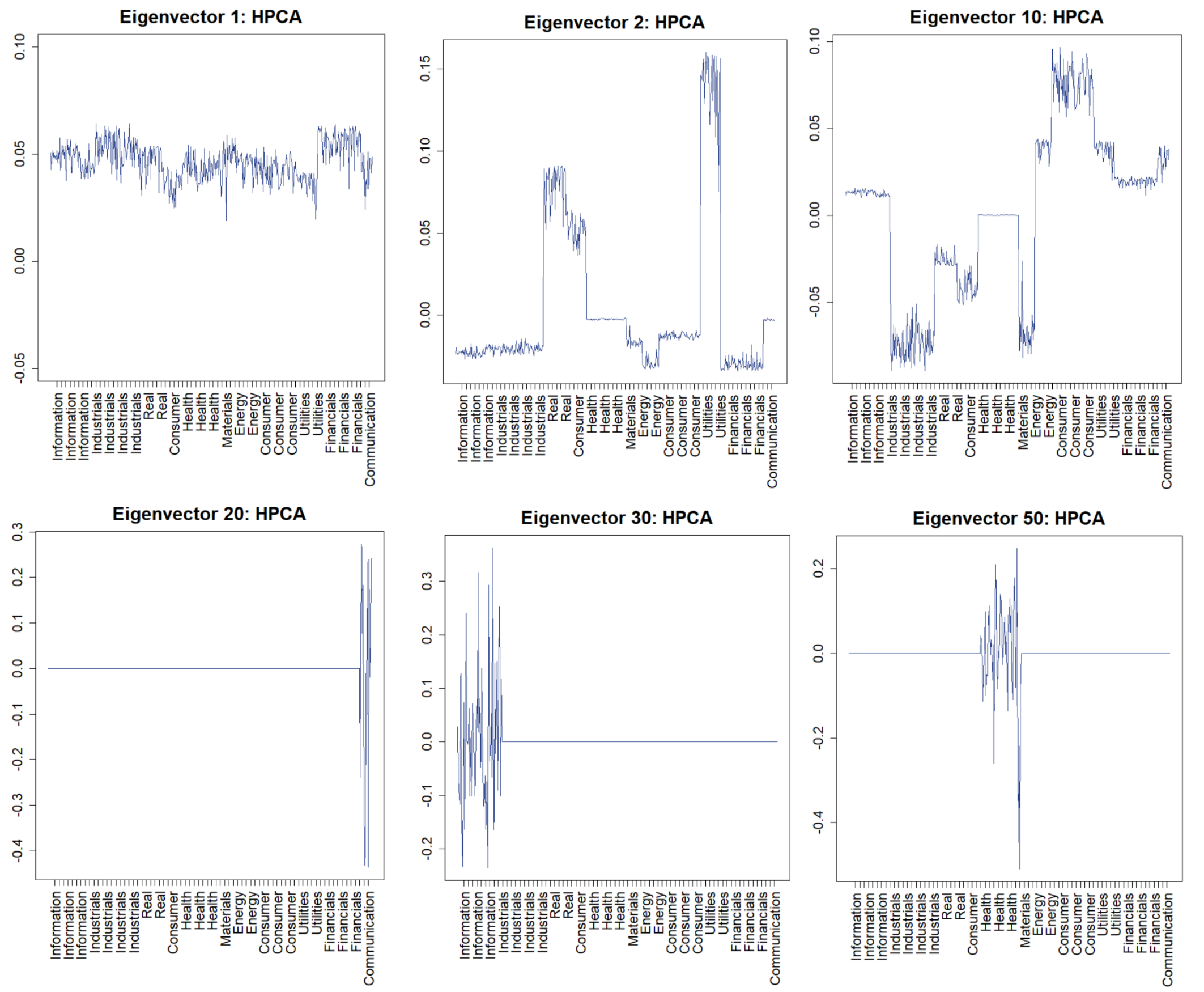}
\caption{\label{USAHPCA} Higher-order eigenvectors of HPCA for US stocks clustered by GICS. Higher-order HPCA eigenvectors are localized in one or a few a sectors.}
\end{figure}

{}Figure \ref{USAHPCA} shows that the first eigenvector has all the coefficients positive and that the higher-order eigenvectors are concentrated in a narrow range of components, which represents specific sectors or group of sectors.\footnote{We took a few eigenvectors to demonstrate how HPCA works on the data.} 

{}Based on this analysis, HPCA can be used to clean the correlation matrix, build corresponding factor models and apply them to portfolio management. The technique mitigates the identification problem by associating higher-order eigevectors to a specific cluster or a group of clusters.

\subsection{China}

{}We took the stock universe to be the constituents of the CSI 500 index. The main GICS  sectors Industrials, Information Technology, Consumer Discretionary and Materials, accounting for more than 70\% of the names.

\subsubsection{Eigenvalue Analysis: clusters based on GICS} 

{} As expected, the PCA explained variance chart rises faster than HPCA. The difference between the two curves is more pronounced here than in the US stocks.\\

\begin{figure}[H]
\includegraphics[width=1.\linewidth]{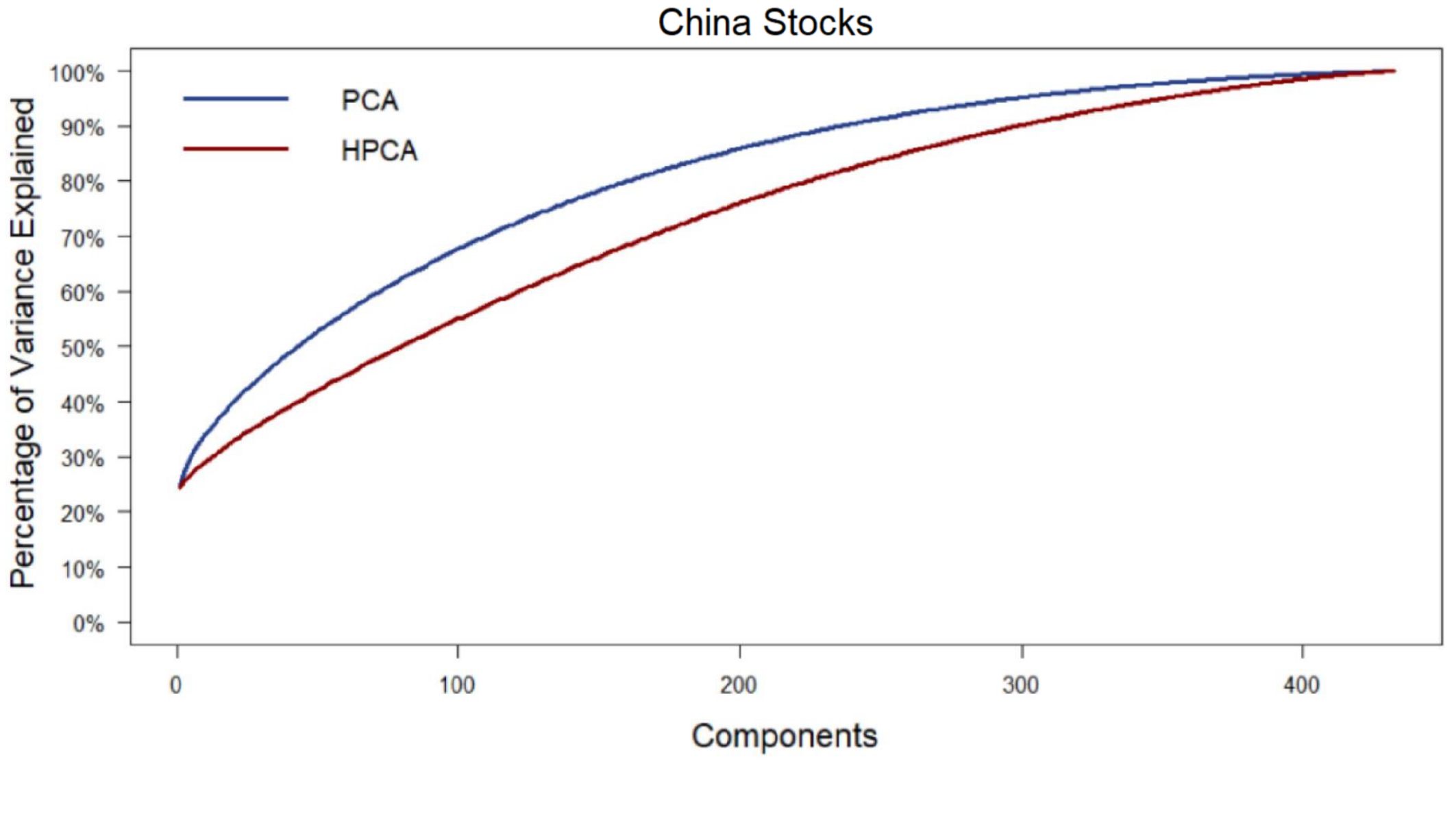}
\caption{Cumulative variance explained by eigenvalues of PCA and HPCA.}
\end{figure}

{}Table \ref{DataChina} shows the same as above. HPCA has lower eigenvalues and the difference in the variance explained by each eigenvalue is substantially higher than in the case of the US stocks.

\begin{table}[H]							
\centering							
\addtolength{\tabcolsep}{-3pt} 
\begin{tabular}[t]{clccc|ccc|llc}						
   \hline						
   \multicolumn{6}{r}{Chinese Stocks}\\ 
   \hline						
  \hline						
&Eigen	&&PCA	&HPCA	&PCA (\%) &HPCA (\%) &&& Eigenportfolio\\  \hline	
  \hline							
& Eigen 1	&&106.03	&105.45	&24.54\%	&24.41\% &&& Multi-sector\\
& Eigen 2	&&7.44	    &3.43	&1.72\%	&0.79\% &&& Multi-sector\\			
& Eigen 3	&&5.94	    &2.32	&1.38\%	&0.54\% &&& Multi-sector\\			
& Eigen 4	&&5.29	    &2.09	&1.22\%	&0.48\% &&& Multi-sector\\			
& Eigen 5	&&4.58	    &2.05	&1.06\%	&0.47\% &&& Multi-sector\\			
& Eigen 6	&&4.08	    &1.99	&0.94\%	&0.46\% &&& Multi-sector\\			
& Eigen 7	&&3.57	    &1.97	&0.83\%	&0.46\% &&& Multi-sector\\			
& Eigen 8	&&3.27	    &1.96	&0.76\%	&0.45\% &&& Multi-sector\\			
& Eigen 9	&&3.09	    &1.87	&0.72\%	&0.43\% &&& Multi-sector\\			
& Eigen 10	&&2.92	    &1.86	&0.68\%	&0.43\% &&& Multi-sector\\			
& Eigen 11	&&2.84	    &1.76	&0.66\%	&0.41\% &&&  Multi-sector\\			
& Eigen 12	&&2.79	    &1.73	&0.65\%	&0.40\% &&& Multi-sector\\			
& Eigen 13	&&2.67	    &1.72	&0.62\%	&0.40\% &&& Multi-sector\\			
& Eigen 14	&&2.62	    &1.70	&0.61\%	&0.39\% &&& Multi-sector\\			
& Eigen 15	&&2.57	    &1.67	&0.59\%	&0.39\% &&& Multi-sector\\			
							
\hline							
\hline							
\end{tabular}							
\caption{\label{DataChina} First 15 HPCA and PCA eigenvalues clustered by GICS for Chinese markets.}							
\end{table}

\subsubsection{Eigenvector Analysis: clusters based on GICS}

\begin{figure}[H]
\includegraphics[width=1\linewidth]{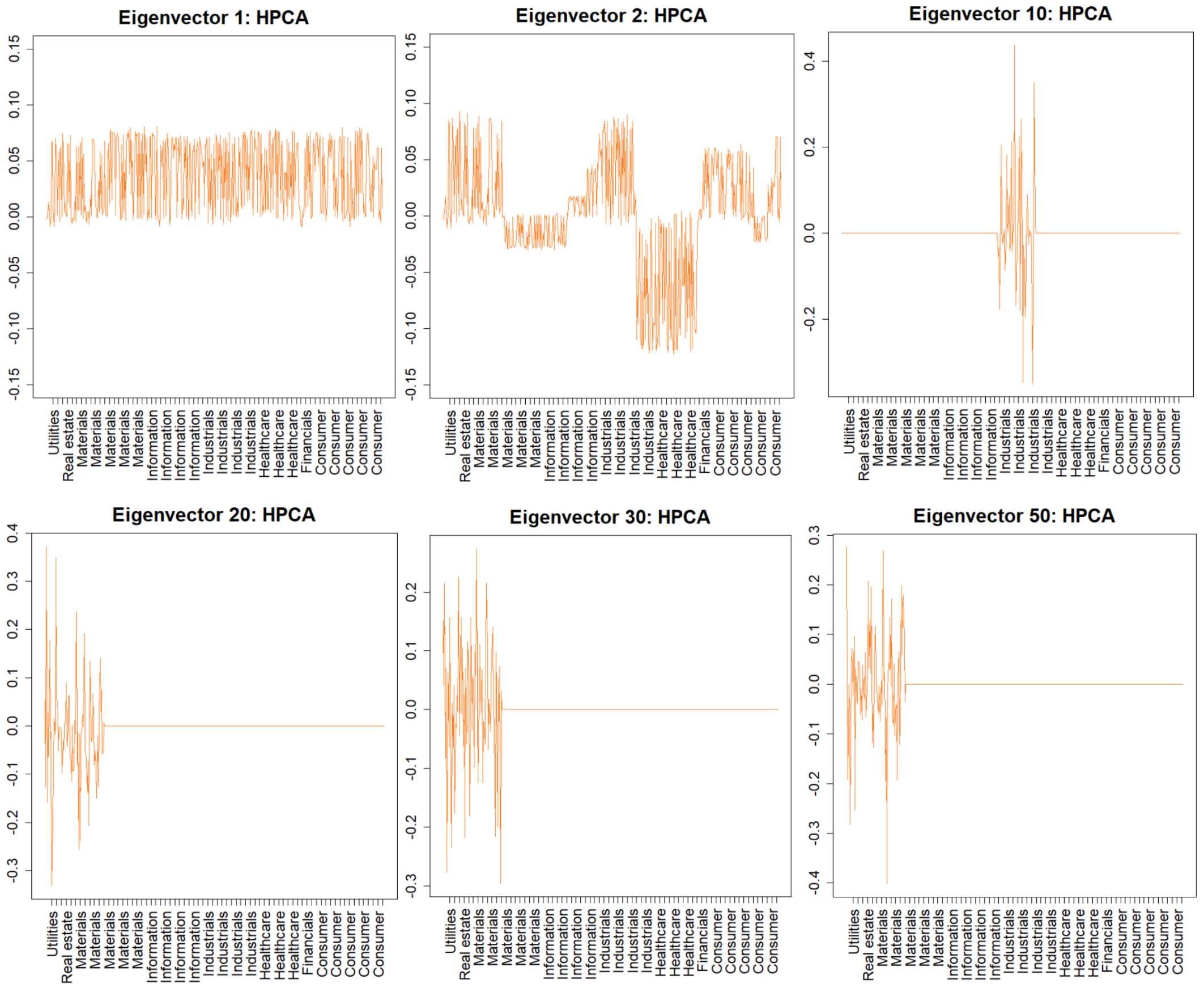}
\caption{\label{HPCACH} Higher-order eigenvectors of HPCA for China markets clustered by GICS. Higher-order HPCA eigenvectors are localized in one or a few sectors.}
\end{figure}

{}Figure \ref{HPCACH} shows different eigenvectors for the case of China. Again, the first eigenvector has all the positive coefficients, representing a good proxy for the market portfolio. Higher-order eigenvectors are concentrated in a narrow range of components, representing specific sectors or group of sectors (i.e., portfolios of eigenportolios), as depicted in Table \ref{DataChina}.

\subsection{European Stocks}

{}In this subsection, we repeat the previous analysis for European stocks, but here, in addition to the GICS clustering, we extend the analysis to clusters based on European countries.

{}The stocks analyzed belongs to the STOXX Europe 600 index, one of the main stock indexes of Europe. The main GICS are Industrials, Financials, Consumer Discretionar and Health Care. The main countries are Great Britain, France, Switzerland, Germany, Netherlands, Sweden, Spain, Italy, among others.

\subsubsection{Eigenvalue Analysis: clusters based on GICS} 

{}As expected, the curve of the PCA cumulative explained variance increases faster than the HPCA curve. The behavior here is similar to that of the US market. \\

\begin{figure}[H]
\includegraphics[width=1.\linewidth]{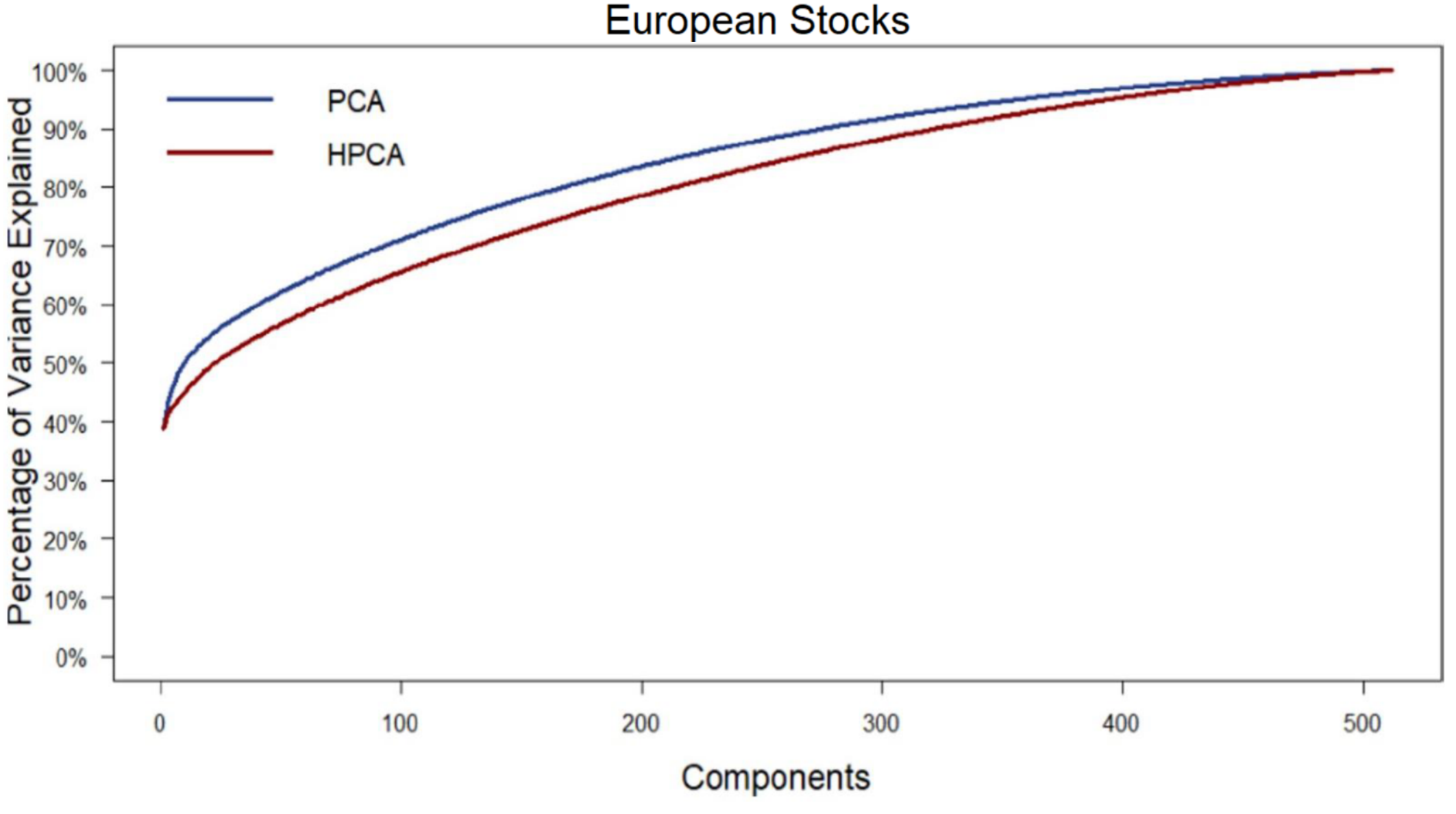}
\caption{Cumulative variance explained by eigenvalues of PCA and HPCA.}
\end{figure}

{}Table \ref{DataEUR} shows that HPCA has slightly lower eigenvalues. We note that the difference between PCA and HPCA is smaller here than in the Chinese stocks, as the market is more diverse.

\begin{table}[H]					
\centering					
\addtolength{\tabcolsep}{-3pt} 
\begin{tabular}[t]{clccc|ccc|llc}						
   \hline						
   \multicolumn{6}{r}{European Stocks}\\ 
   \hline						
  \hline						
&Eigen	&&PCA	&HPCA	&PCA (\%) &HPCA (\%) &&& Eigenportfolio\\		
  \hline					
& Eigen 1	&&197.67	&194.94	&38.76\%	&38.20\%	&&&Multi-sector\\	
& Eigen 2	&&11.42		&7.45	&2.24\%		&1.46\%	&&&Multi-sector\\	
& Eigen 3	&&10.13		&5.82	&1.99\%		&1.14\%	&&&Multi-sector\\	
& Eigen 4	&&9.24		&3.96	&1.81\%		&0.78\%	&&&Multi-sector\\	
& Eigen 5	&&6.28		&3.18	&1.23\%		&0.63\%	&&&Financials\\	
& Eigen 6	&&5.63		&2.94	&1.10\%		&0.58\%	&&&Multi-sector\\	
& Eigen 7	&&4.80		&2.66	&0.94\%		&0.52\%	&&&Consumer Disc.\\
& Eigen 8	&&3.88		&2.54	&0.76\%		&0.50\%	&&&Industrials\\	
& Eigen 9	&&3.64		&2.47	&0.71\%		&0.48\%	&&&Financials\\	
& Eigen 10	&&3.40		&2.43	&0.67\%		&0.48\%	&&&Multi-sector\\	
& Eigen 11	&&3.05		&2.28	&0.60\%		&0.45\%	&&&Multi-sector\\	
& Eigen 12	&&2.68		&2.26	&0.53\%		&0.44\%	&&&Multi-sector\\	
& Eigen 13	&&2.49		&2.17	&0.49\%		&0.43\%	&&&Materials\\	
& Eigen 14	&&2.37		&2.16	&0.47\%		&0.42\%	&&&Multi-sector\\	
& Eigen 15	&&2.14		&2.12	&0.42\%		&0.41\%	&&&Multi-sector\\
\hline					
\hline					
\end{tabular}					
\caption{\label{DataEUR} First 15 HPCA and PCA eigenvalues clustered by GICS for the European market.}					
\end{table}					

\subsubsection{Eigenvector Analysis: clusters based on GICS}

\begin{figure}[H]
\includegraphics[width=1\linewidth]{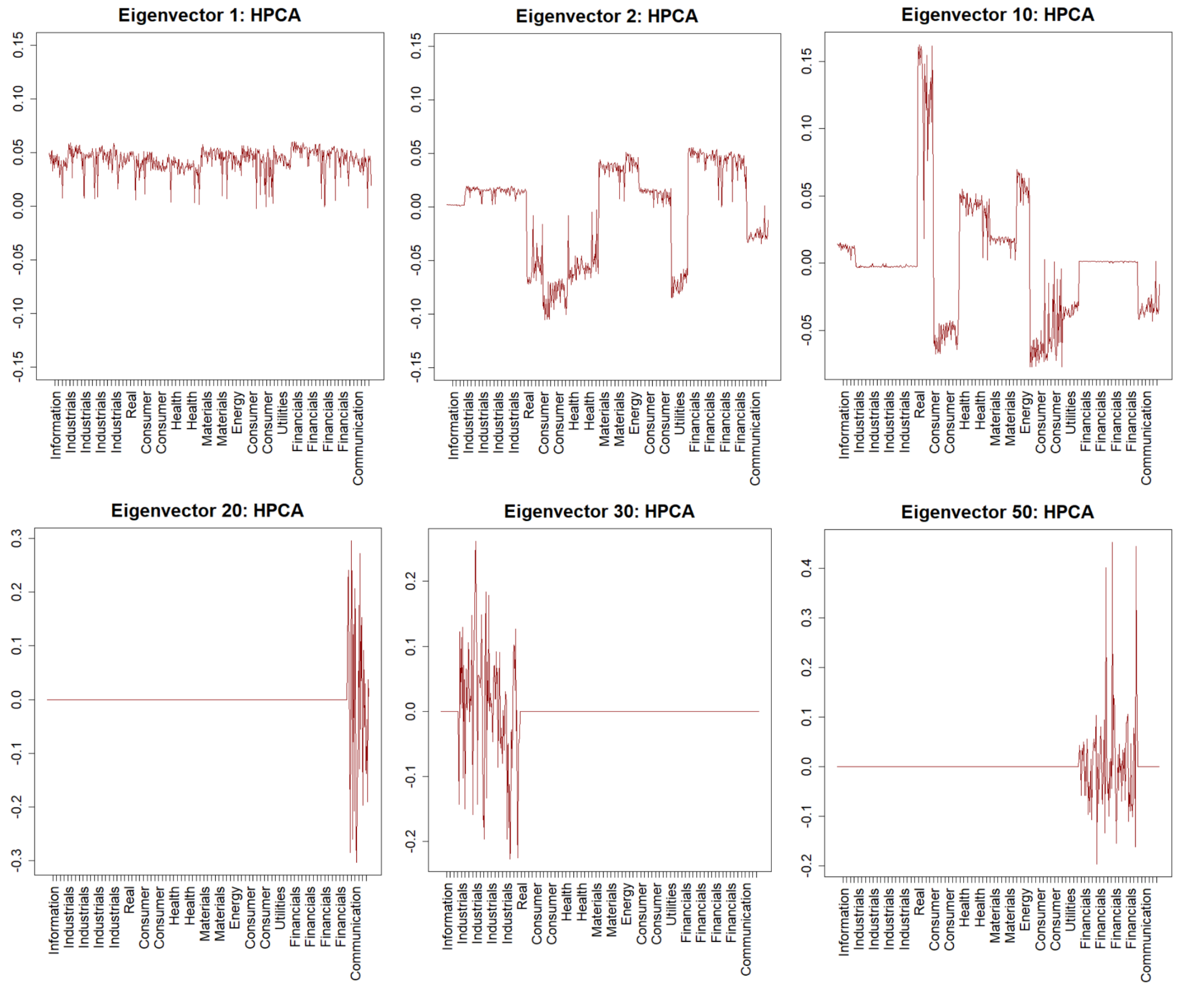}
\caption{\label{HPCAEU} Higher-order eigenvectors of HPCA for European markets clustered by GICS. Higher-order HPCA eigenvectors are localized in one or a few sectors.}
\end{figure}

{}Figure \ref{HPCAEU} shows that the first eigenvector represents a long-only portfolio. Higher-order eigenvectors are concentrated in a narrow range of components.

{}For example, the twentieth eigenportfolio is concentrated with long and short coefficients in the Communication and Financial sectors. The thirteenth eigenportfolio focuses on Industrial and Real Estate and the fifteenth has its coefficients almost entirely in the Financial sector. The second and the tenth are less intuitive, with significant coefficients across the spectrum.  Table \ref{DataEUR} shows which components represent specific sectors or groups of sectors.

\subsubsection{Eigenvalue Analysis: clusters based on Countries}

{}Here we conducted HPCA based on the main countries belonging to the European market (see Table \ref{DataCountries} in the Appendix for more details). As in the GICS-based analysis, the curve of cumulative explained variance of PCA increases faster than the HPCA counterpart. In addition, compared to HPCA for the European stocks based on GICS, here the first eigenvalues explain a greater variation, meaning a higher level of concentration in a few components. For example, the first GICS-based eigenvalue of HPCA explains 33.36\% of the total variance, while the country-based first eigenvalue of HPCA explains 38.76\%. 

\begin{figure}[H]
\includegraphics[width=1.\linewidth]{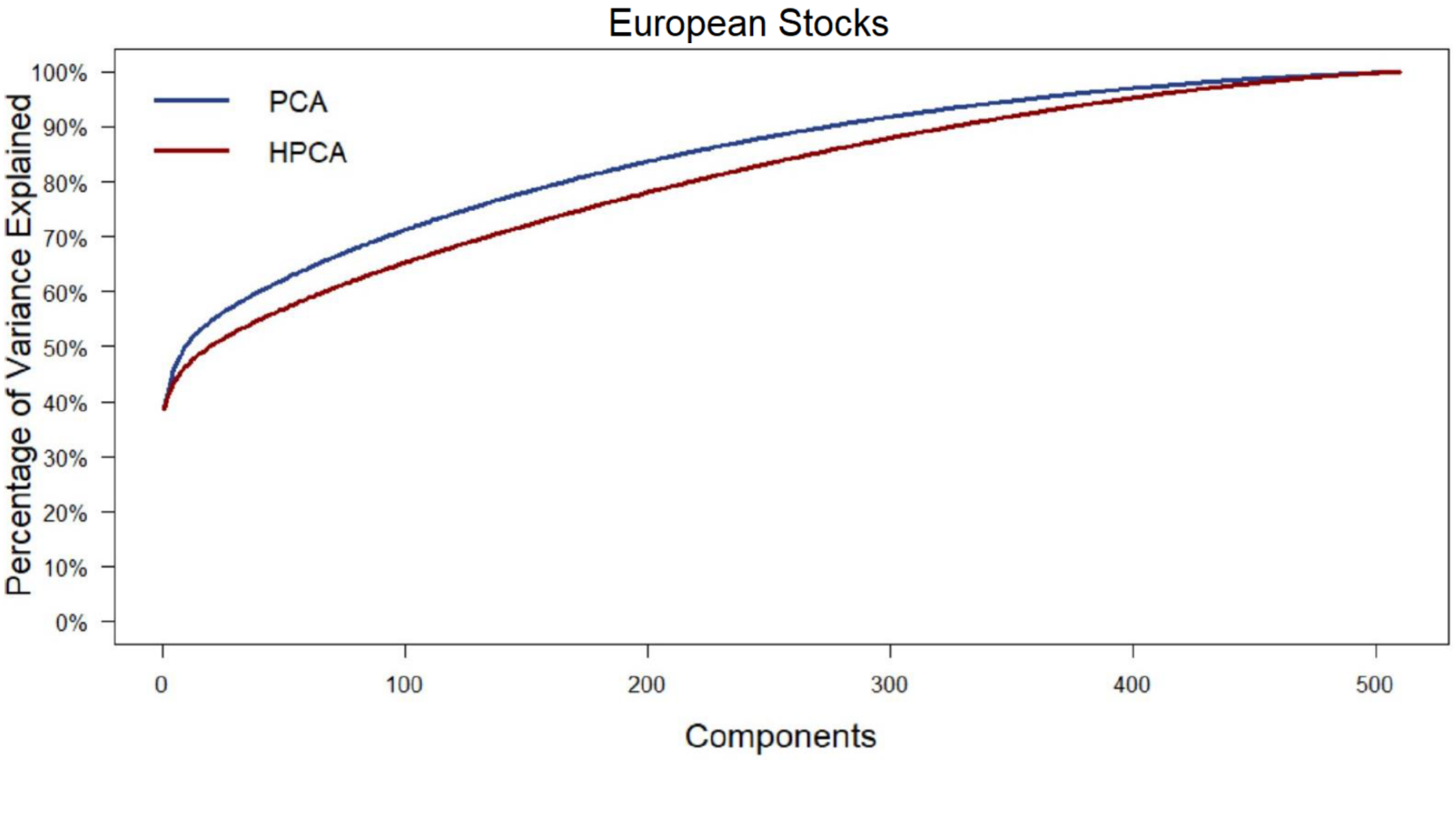}
\caption{Cumulative variance explained by eigenvalues of PCA and HPCA based on country clustering.}
\end{figure}

{}Table \ref{DataEUR1} shows that HPCA has slightly lower eigenvalues than PCA. As mentioned earlier, the level concentration is higher than in the case of GICS-based HPCA. Also, here the difference between PCA and HPCA is smaller. Some components represent multiple countries (portfolios of eigenportfolios) while others are concentrated in only one country. For example, the sixth and the seventh eigenportfolios are concentrated in United Kingdom, while the thirteenth and fourteenth represent Switzerland and Germany, respectively.  

\begin{table}[H]					
\centering					
\addtolength{\tabcolsep}{-3pt} 
\begin{tabular}[t]{clccc|ccc|llc}						
   \hline						
   \multicolumn{6}{r}{European Stocks}\\ 
   \hline						
  \hline						
&Eigen	&&PCA	&HPCA	&PCA (\%) &HPCA (\%) &&& Eigenportfolio\\\hline		
& Eigen 1	&&197.67 &197.18 &38.76\% &38.66\% &&& Multi-country\\	
& Eigen 2	&&11.42	&9.14	&2.24\%	&1.79\%  &&& Multi-country\\	
& Eigen 3	&&10.13	&5.65	&1.99\%	&1.11\%  &&& Multi-country\\	
& Eigen 4	&&9.24	&5.39	&1.81\%	&1.06\%  &&& Multi-country\\	
& Eigen 5	&&6.28	&4.33	&1.23\%	&0.85\% &&& Multi-country\\	
& Eigen 6	&&5.63	&3.86	&1.10\%	&0.76\%  &&& United Kingdom \\	
& Eigen 7	&&4.80	&3.36	&0.94\%	&0.66\% &&&  United Kingdom\\	
& Eigen 8	&&3.88	&2.78	&0.76\%	&0.54\% &&&  Multi-country\\	
& Eigen 9	&&3.64	&2.74	&0.71\%	&0.54\% &&&  Multi-country\\	
& Eigen 10	&&3.40	&2.48	&0.67\%	&0.49\% &&&  Multi-country\\	
& Eigen 11	&&3.05	&2.19	&0.60\%	&0.43\% &&& Multi-country\\	
& Eigen 12	&&2.68	&2.16	&0.53\%	&0.42\% &&&  Multi-country\\	
& Eigen 13	&&2.49	&2.13	&0.49\%	&0.42\% &&&  Switzerland\\	
& Eigen 14	&&2.37	&1.96	&0.47\%	&0.38\% &&& Germany\\	
& Eigen 15	&&2.14	&1.86	&0.42\%	&0.37\%&&&  Multi-country\\	
					
\hline					
\hline					
\end{tabular}					
\caption{\label{DataEUR1} First 15 HPCA and PCA eigenvalues clustered by country for European markets.}					
\end{table}

\subsubsection{Eigenvector Analysis: clusters based on Countries}

{}The eigenvectors here are unequivocal, showing the power of HPCA. The first eigenvector represents a market portfolio across European countries. Higher-order portfolios are concentrated in a few countries. For example, the tenth, the twentieth and the thirtieth portfolios are the three a combination of France and Germany. The fifteenth eigenportfolio is concentrated with almost all the positive coefficients in Norway and Finland. See Table \ref{DataEUR1} for more details.

\begin{figure}[H]
\includegraphics[width=1\linewidth]{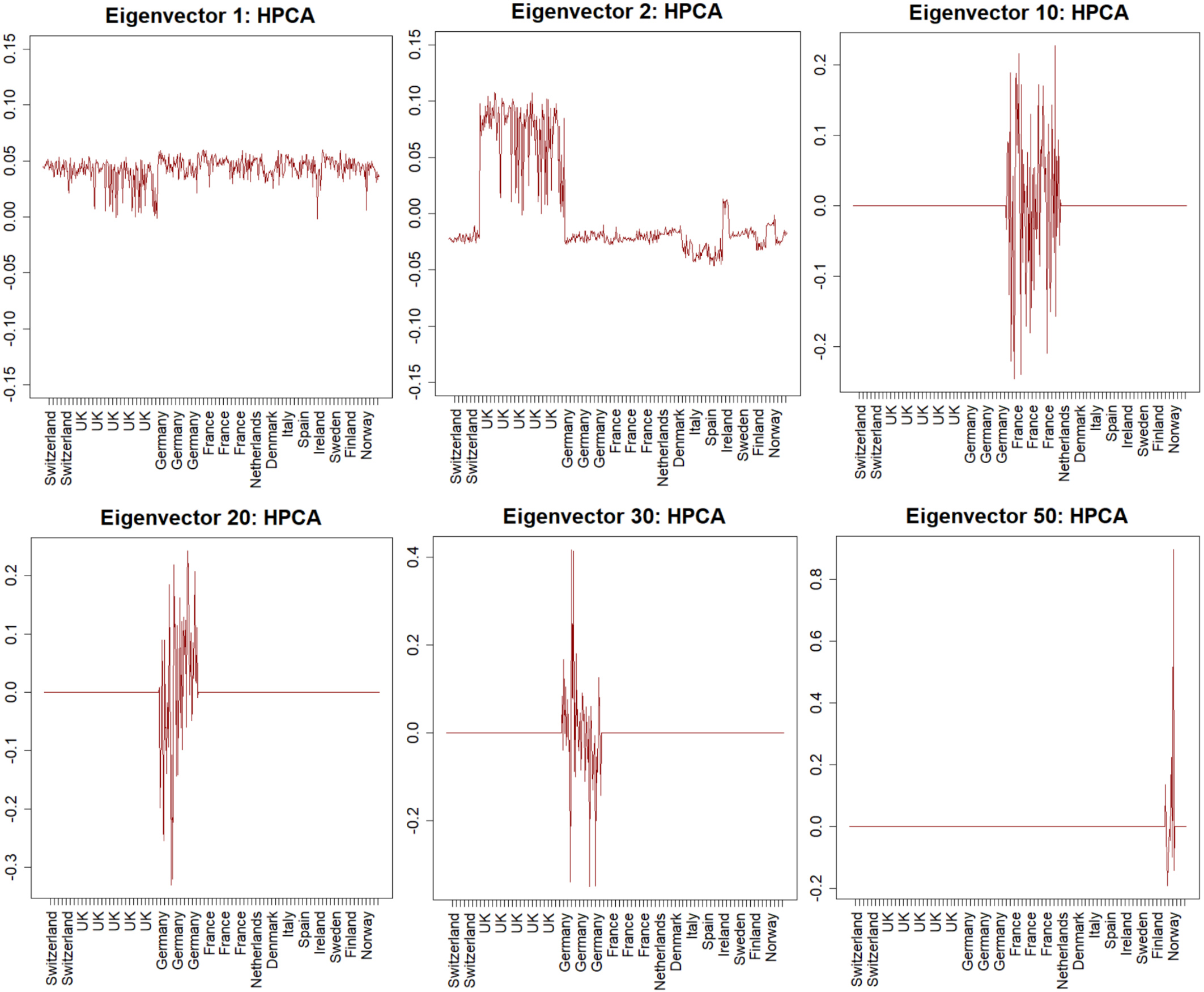}
\caption{\label{HPCAEUR} Higher-order eigenvectors of HPCA for European markets clustered by countries. Higher-order HPCA eigenvectors are localized in one or a few countries.}
\end{figure}

\subsection{Emerging Markets}

{}As in the case of European stocks, here we cluster the stocks based on GICS and countries. The stocks analyzed belongs to the MSCI Emerging Markets Index. Almost 60\% of the stocks are concentrated in Financials, Information Technology, Materials, and Consumer Discretionary. The main countries are China, Korea, Taiwan, India, Brazil, South Africa, Russia, Mexico, and Thailand.

\subsubsection{Eigenvalue Analysis: clusters based on GICS} 

{}The cumulative explained variance curve of PCA increases faster than the HPCA curve. The behavior here is similar to that of the Chinese market, i.e., the difference in the explained variance between PCA and HPCA is wider than in the case of the US and the European markets.

\begin{figure}[H]
\includegraphics[width=1.\linewidth]{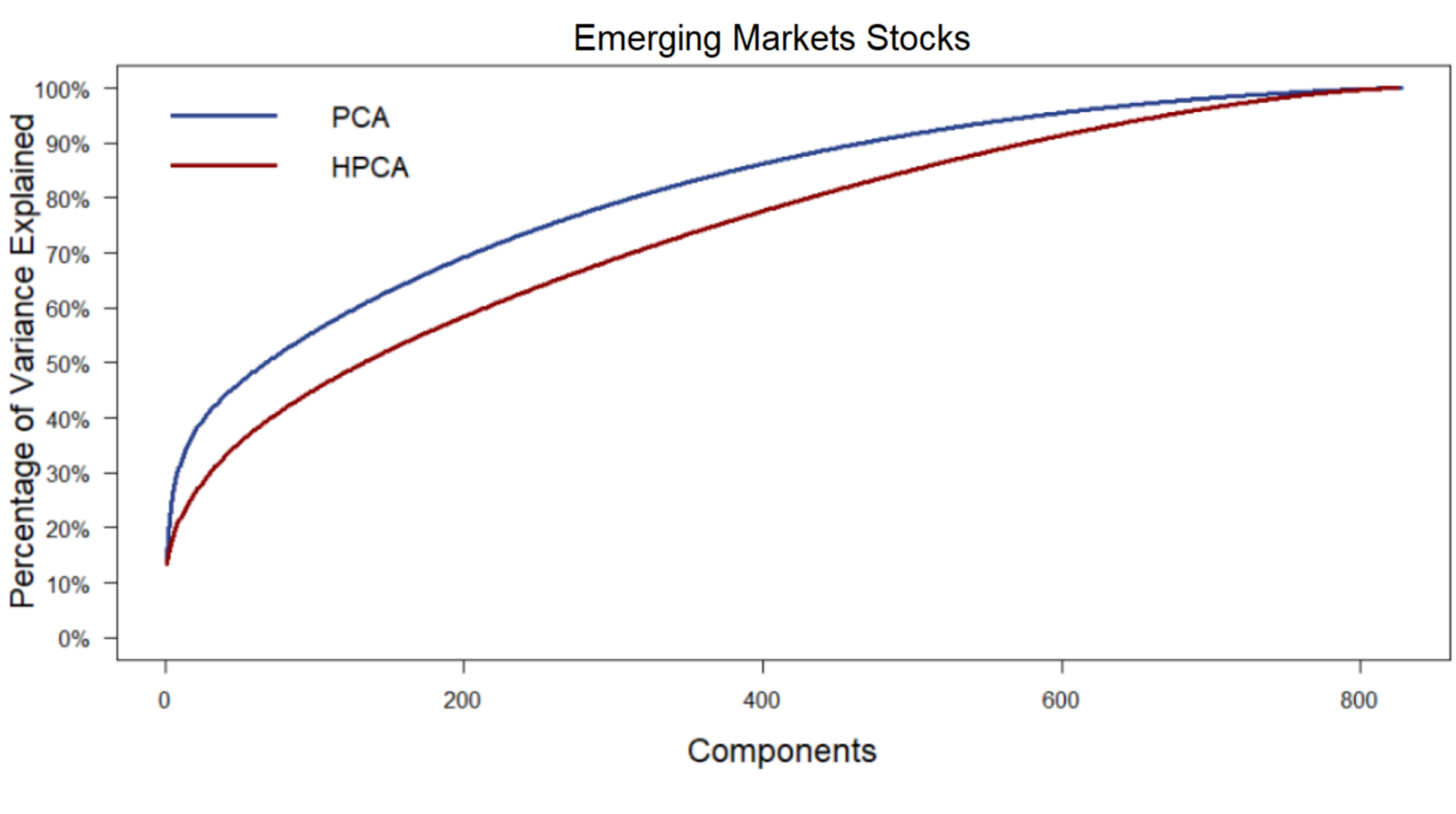}
\caption{Cumulative variance explained by eigenvalues of PCA and HPCA.}
\end{figure}

{}Table \ref{DataEME} shows that HPCA has lower eigenvalues than PCA, repeating the pattern observed in all the previous cases analyzed. Furthermore, the first eigenvalue explains significantly less variance than in all other cases. To put it in perspective, the first eigenvalue of PCA and HPCA accounts approximately for 13\% of the total variance, whereas in the US, Chinese and European markets represent approximately 37\%, 24\% and 33\% of the total variance, respectively.

\begin{table}[H]					
\centering					
\addtolength{\tabcolsep}{-3pt} 
\begin{tabular}[t]{clccc|ccc|llc}						
   \hline						
   \multicolumn{6}{r}{Emerging Stocks}\\ 
   \hline						
  \hline						
&Eigen	&&PCA	&HPCA	&PCA (\%) &HPCA (\%) &&& Eigenportfolio\\  \hline
& Eigen 1	&&113.67 &109.19 &13.83\%	&13.24\%	&&& Multi-sector\\
& Eigen 2	&&39.44	&14.72	&4.80\%	&1.78\%	&&&Financials\\
& Eigen 3	&&23.77	&8.98   &2.89\%	&1.09\%	&&&Multi-sector\\
& Eigen 4	&&20.68	&8.10	&2.52\%	&0.98\%	&&&Multi-sector\\
& Eigen 5	&&14.87	&6.97	&1.81\%	&0.85\%	&&&Inf. Technology\\
& Eigen 6	&&11.27	&6.95	&1.37\%	&0.84\%	&&&Multi-sector\\
& Eigen 7	&&10.81	&6.63	&1.32\%	&0.80\%	&&&Multi-sector\\
& Eigen 8	&&9.30	&6.19	&1.13\%	&0.75\%	&&&Multi-sector\\
& Eigen 9	&&8.68	&5.75	&1.06\%	&0.70\%	&&&Industrials\\
& Eigen 10	&&7.14	&4.50	&0.87\%	&0.55\%	&&&Consumer Stap.\\
& Eigen 11	&&6.35	&4.38	&0.77\%	&0.53\%	&&&Financials\\
& Eigen 12	&&6.07	&4.33	&0.74\%	&0.53\%	&&&Financials\\
& Eigen 13	&&5.94	&4.21	&0.72\%	&0.51\%	&&&Consumer Stap.\\
& Eigen 14	&&5.37	&3.89	&0.65\%	&0.47\%	&&&Multi-sector\\
& Eigen 15	&&4.81	&3.87	&0.59\%	&0.47\%	&&&Multi-sector\\
\hline									
\hline									
\end{tabular}									
\caption{\label{DataEME} First 15 HPCA and PCA eigenvalues clustered by GICS for Emerging markets.}									
\end{table}

\subsubsection{Eigenvector Analysis: clusters based on GICS}

\begin{figure}[H]
\includegraphics[width=1\linewidth]{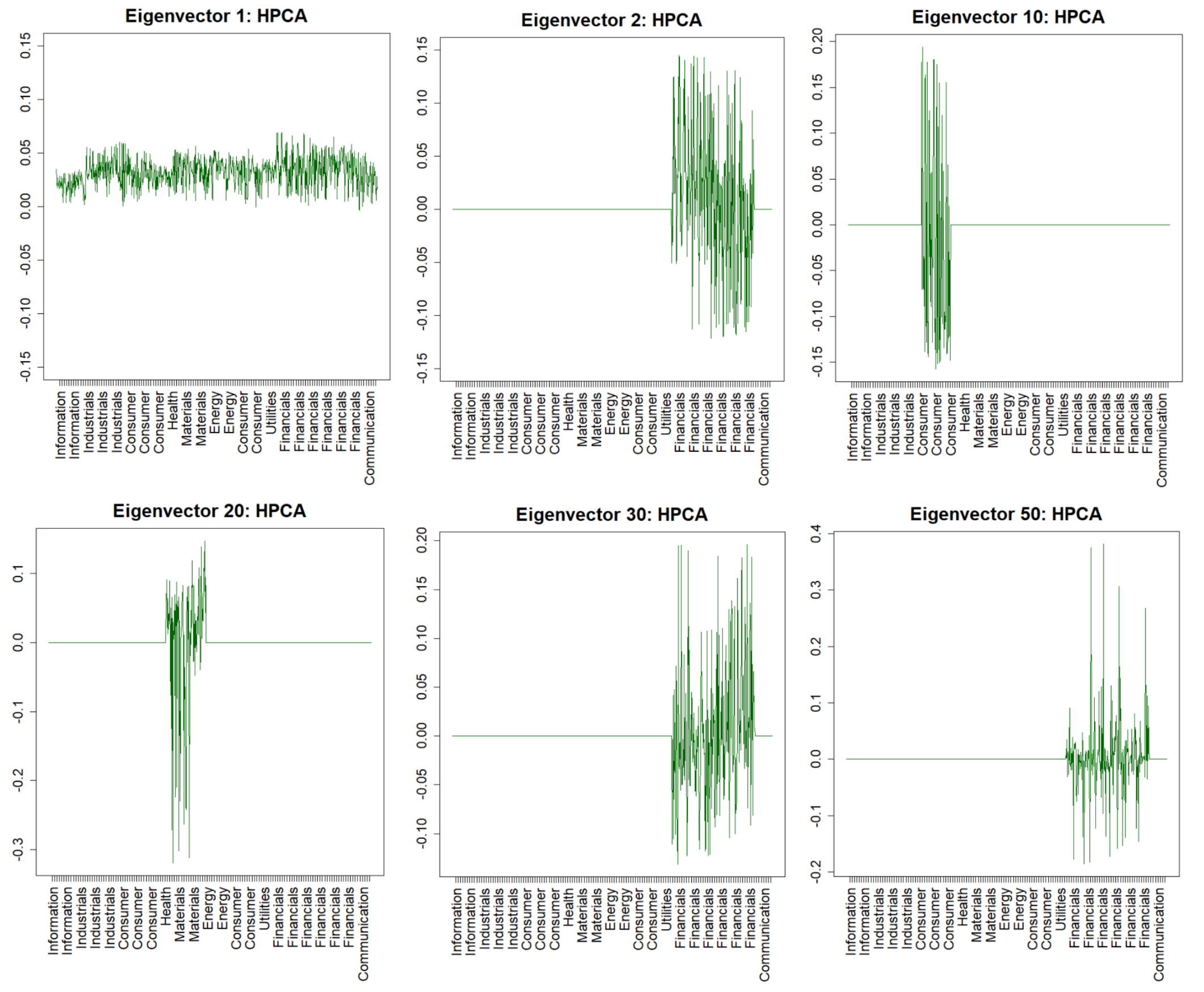}
\caption{\label{HPCAEM} Higher-order eigenvectors of HPCA for Emerging markets clustered by GICS. Higher-order HPCA eigenvectors are localized in one or a few sectors.}
\end{figure}

{}Figure \ref{HPCAEM} shows that the first eigenvector represents a proxy for a market portfolio. Higher-order eigenvectors are concentrated in a narrow range of components.

{}For example, the second eigenvector is concentrated exclusively in the Financial sector, the tenth in Consumer Staples, the twentieth eigenportfolio is concentrated with long and short coefficients in the Material and Energy sectors. The thirteenth eigenportfolio focuses on Financial and Utilities sectors and the fifteenth has its coefficients almost entirely in the Financial sector. See Table \ref{DataEME} for more details on portfolios.

\subsubsection{Eigenvalue Analysis: clusters based on Countries}

{}Like all the previous cases, the curve of the PCA cumulative explained variance increases faster than the HPCA curve. The behavior here is slightly different that the behavior of the GICS-based HPCA for Emerging markets. Specifically, the different of the curves of the two approaches (PCA and HPCA) here is lower than the counterpart for the GICS-based HPCA case.

\begin{figure}[H]
\includegraphics[width=1.\linewidth]{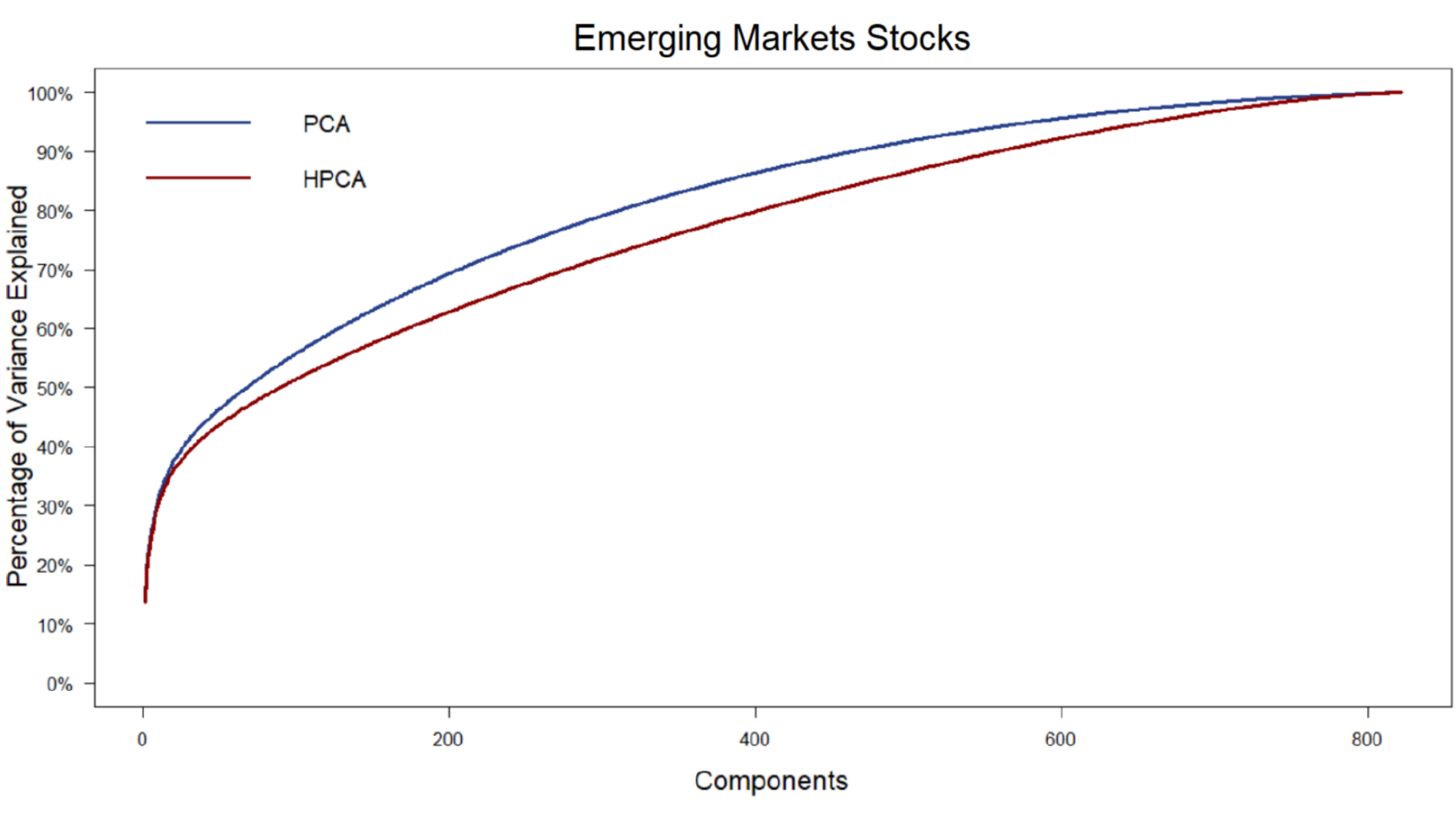}
\caption{Cumulative variance explained by eigenvalues of PCA and HPCA classified by country.}
\end{figure}

{}Table \ref{DataEME1} shows that HPCA has slightly lower eigenvalues. Furthermore, it shows that almost all eigenportfolios are built based on multiple countries.

\begin{table}[H]							
\centering							
\addtolength{\tabcolsep}{-3pt} 
\begin{tabular}[t]{clccc|ccc|llc}						
   \hline						
   \multicolumn{6}{r}{Emerging Stocks}\\ 
   \hline						
  \hline						
&Eigen	&&PCA	&HPCA	&PCA (\%) &HPCA (\%) &&& Eigenportfolio\\  \hline	
  \hline							
& Eigen 1	&&113.67 &111.51 &13.83\% &13.57\% &&& Multi-country\\
& Eigen 2	&&39.44	&38.20	&4.80\%	&4.65\% &&& Multi-country\\
& Eigen 3	&&23.77	&22.03	&2.89\%	&2.68\% &&& Multi-country\\
& Eigen 4	&&20.68	&19.00	&2.52\%	&2.31\% &&& Multi-country\\
& Eigen 5	&&14.87	&13.93	&1.81\%	&1.69\% &&& Multi-country\\
& Eigen 6	&&11.27	&11.33	&1.37\%	&1.38\% &&& China\\
& Eigen 7	&&10.81	&10.85	&1.32\%	&1.32\% &&& Multi-country\\
& Eigen 8	&&9.30	&9.66	&1.13\%	&1.18\% &&& Multi-country\\
& Eigen 9	&&8.68	&8.22	&1.06\%	&1.00\% &&& Multi-country\\
& Eigen 10	&&7.14	&6.88	&0.87\%	&0.84\% &&& Multi-country\\
& Eigen 11	&&6.35	&5.92	&0.77\%	&0.72\% &&& Multi-country\\
& Eigen 12	&&6.07	&5.66	&0.74\%	&0.69\% &&& China\\
& Eigen 13	&&5.94	&5.42	&0.72\%	&0.66\% &&& Multi-country\\
& Eigen 14	&&5.37	&4.62	&0.65\%	&0.56\% &&& Multi-country\\
& Eigen 15	&&4.81	&4.45	&0.59\%	&0.54\% &&& Multi-country\\
\hline							
\hline							
\end{tabular}							
\caption{\label{DataEME1} First 15 HPCA and PCA eigenvalues clustered by country for Emerging markets.}							
\end{table}

\subsubsection{Eigenvector Analysis: clusters based on Countries}

\begin{figure}[H]
\includegraphics[width=1\linewidth]{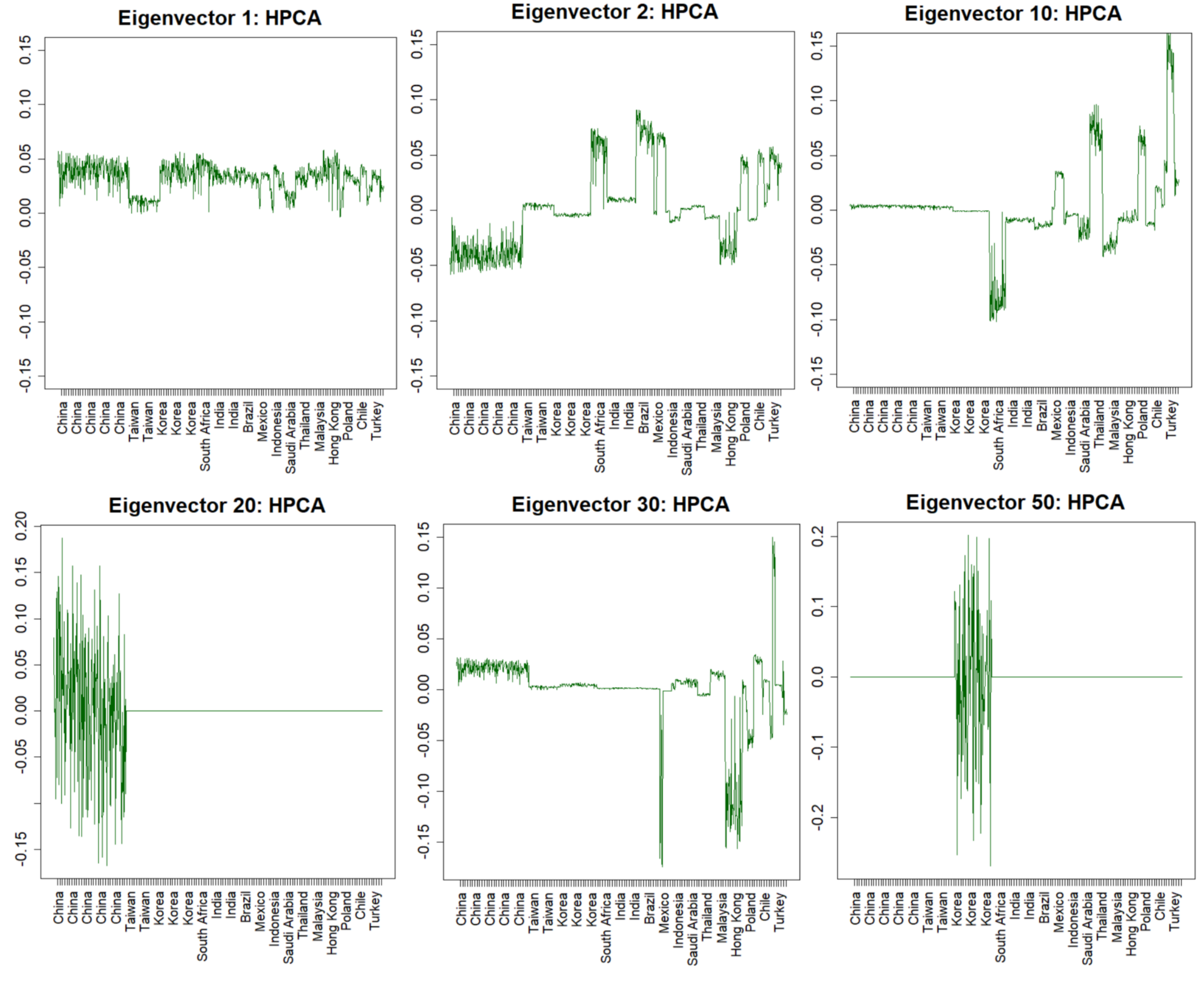}
\caption{\label{HPCAEMcountry} Higher-order eigenvectors of HPCA for Emerging Markets clustered by country.}
\end{figure}

{}Figure \ref{HPCAEMcountry} shows that, unlike the previous GICS-based HPCA case, the distribution of the eigenvectors across the spectrum is less clear here.

{}As usual, the first eigenvector has all the positive coefficients. The other eigenvectors are less intuitive, although it is noticeable that the twentieth eigenportfolio is concentrated almost entirely in China and the fiftieth eigenportfolio in Korea. Table \ref{DataEME1} provides more details. 

\section{HPCA Statistically Generated Clusters}

{}Practitioners usually employ pre-built (static) clusters such as GICS and countries to build factor models. Stocks belonging to the same GICS or country share common factors that capture --to some extend-- their joint dynamics.

{}However, these types of models have some shortcomings. Stock markets and their components change almost continuously, producing substantial changes in their behavior that are not captured by static clusters. This is not desirable for risk and portfolio management for various reasons.

{}It goes against the diversification of a (seemingly diversified) strategy. Many investment portfolios base their mandates on diversifying their allocations among sectors, sub-sectors, countries, etc., to avoid high and undesirable idiosyncratic risk. However, there are several factors beyond the sector and/or the country that affect the behavior of portfolio holdings. For example, when interest rates rise sharply, capital-intensive companies are negatively affected and diversification vanishes.
 
{}Trading strategies, such as the so-called sector/country rotation may also been affected for the same reasons. Securities that belong to a specific sector/country can change their behavior sharply under the changes of a market regime and the strategy that worked ex-ante may stop working overnight.
 
{}To mitigate this problem and account for hidden risk factors, we adopt a purely statistical technique. This is a simple and still powerful tool that dynamically adapts to changes in market conditions over time, which makes it suitable for managing trading portfolios. Also, it is a parsimonious approach since it does not rely on too many parameters. The user only needs to define the number of clusters, which depends on the number of $K$ eigenvectors, without specifying any other parameters or hyper-parameters.

\subsection{Description of the Algorithm}

\begin{figure}[H]
\centering
\includegraphics[width=0.70\linewidth]{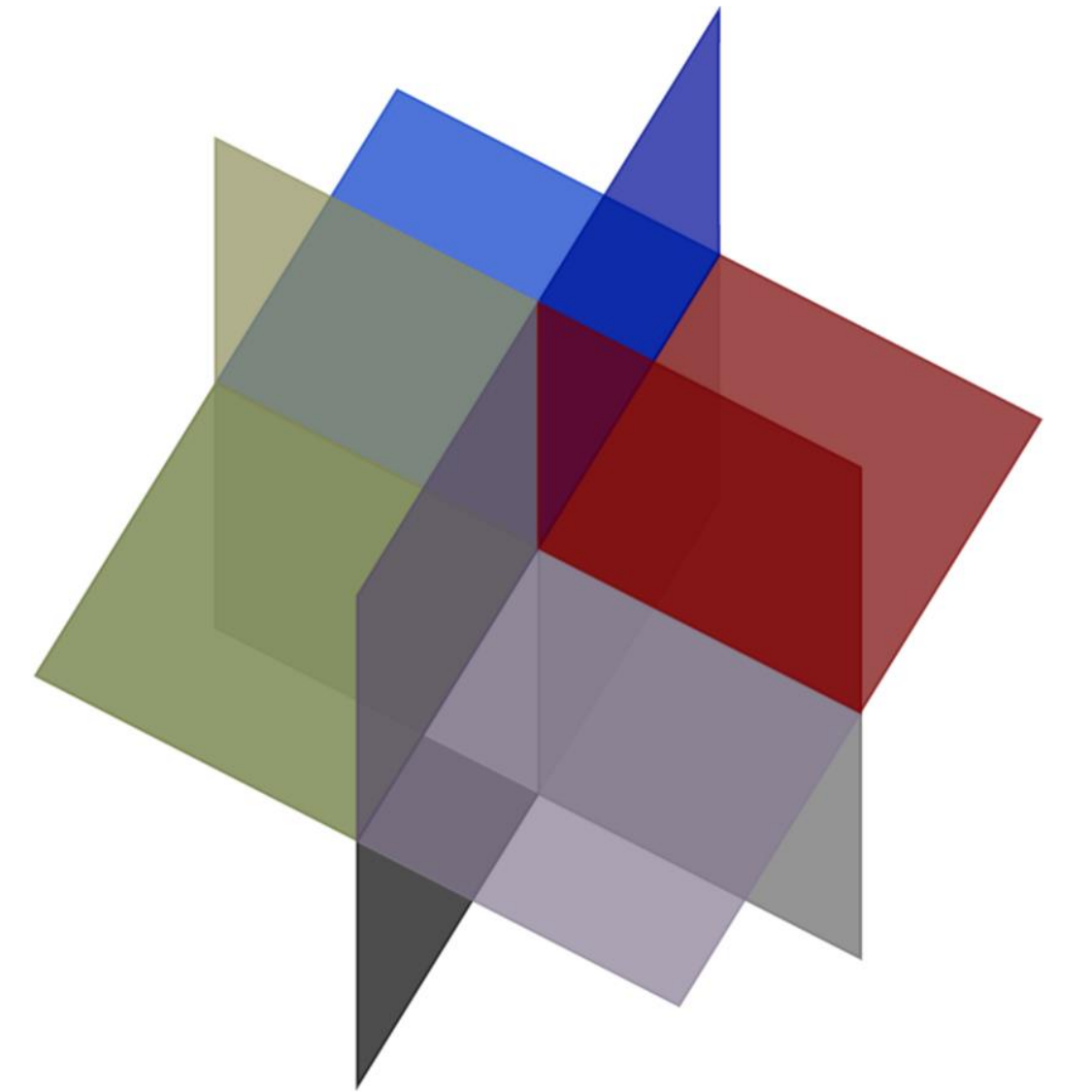}
\caption{\label{quad} The space is divided into different quadrants (clusters) to which each asset belongs based on the sign of the eigenvectors.}
\end{figure}

{}Based on matrix diagonalization (PCA), the algorithm constructs new features in the space that retain the behavior of each component based on linear combinations of its main characteristics, leading to statistical clusters of similar-behaved securities. Figure \ref{quad} illustrates how the space is divided into different quadrants (clusters) to which each stock belongs.

{}To run the statistical clustering, it is needed to estimate the eigenvectors as in Eq. (\ref{Eigen2}) and define the number of factors $K$, omitting the first one (Eq. (\ref{Eigen1})) since it has all the coefficients positive.\footnote{This applies to correlated markets, e.g. stocks. If other (uncorrelated) assets are included, this is not necessary.} Then, each security is clustered appropriately according to the sign of the coefficients in each eigenvector.

\subsection{HPCA with Statistical Clustering}

{}In this section, we analyze HPCA using the statistical clustering method. We set the number of eigevectors $K$ equal to 4. Therefore, the expected number of clusters is $ 2 ^ 4 = $ 16, although not necessarily all the clusters will have components. Some of them can be empty, \textit{ergo} removed.

\subsubsection{Eigenvalue Analysis}

{}Clustering using a statistical approach delivers similar patterns on the curve of cumulative explained variance as the static approaches. The PCA curve rises faster than the HPCA curve.

\begin{figure}[H]
\includegraphics[width=1.\linewidth]{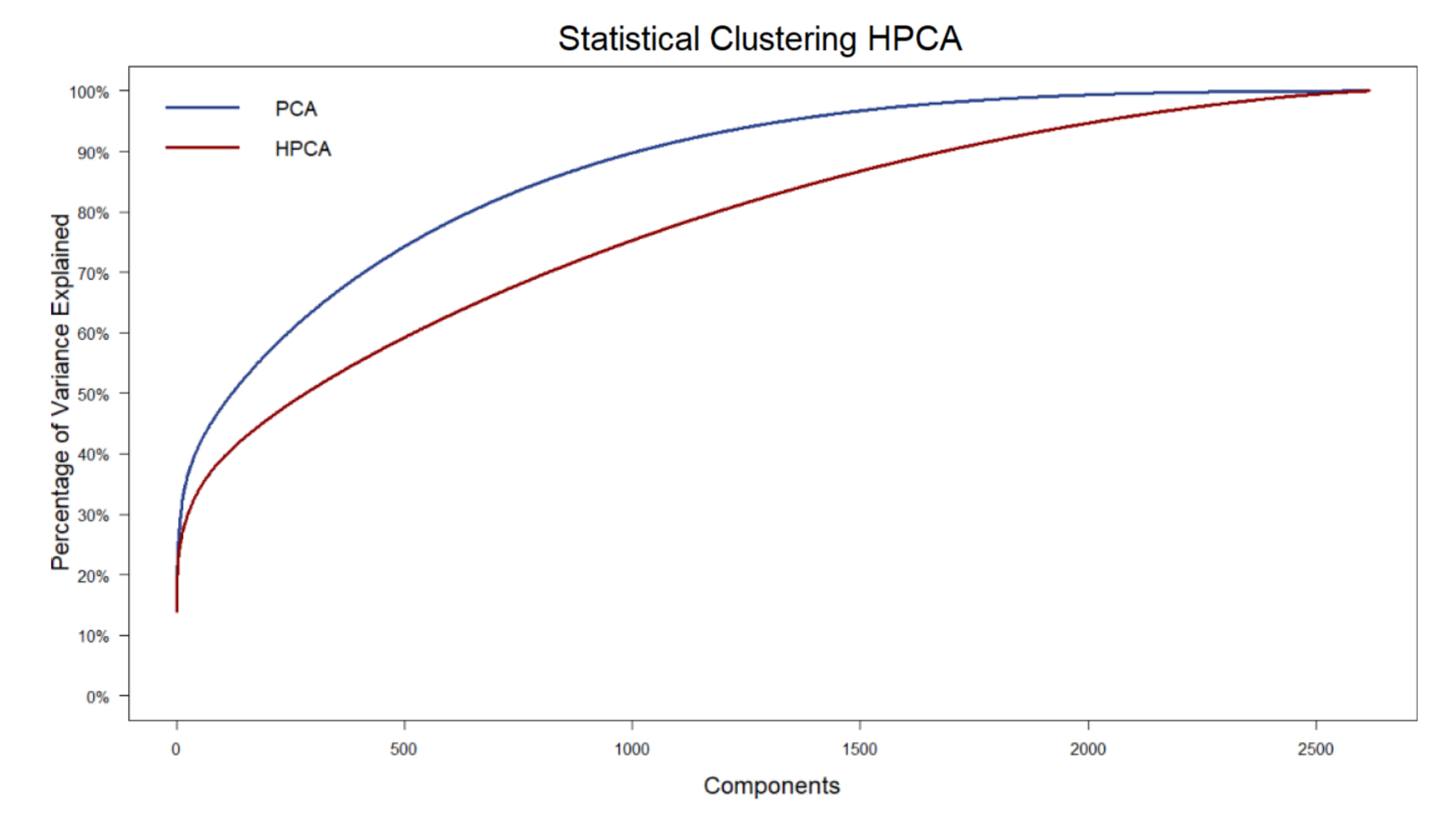}
\caption{Cumulative variance explained by eigenvalues of PCA and HPCA using statistical clustering.}
\end{figure}

{}Table \ref{DataStat} shows that HPCA has lower eigenvalues.

\begin{table}[H]							
\centering							
\addtolength{\tabcolsep}{-3pt} 
\begin{tabular}[t]{clccccc}							
   \hline							
   & & & & Statistical Clustering & & \\							
   \hline							
  \hline							
&Eigen	&&PCA	&HPCA	&PCA (\%)	&HPCA (\%)	\\		
  \hline							
& Eigen 1	&&376.72	&371.51	&14.41\%	&14.21\%\\			
& Eigen 2	&&149.51	&147.68	&5.72\%	&5.65\%\\			
& Eigen 3	&&73.52	&66.99	&2.81\%	&2.56\%\\			
& Eigen 4	&&52.10	&42.09	&1.99\%	&1.61\%\\			
& Eigen 5	&&32.44	&32.17	&1.24\%	&1.23\%\\			
& Eigen 6	&&28.17	&23.85	&1.08\%	&0.91\%\\			
& Eigen 7	&&25.16	&23.36	&0.96\%	&0.89\%\\			
& Eigen 8	&&23.12	&21.52	&0.88\%	&0.82\%\\			
& Eigen 9	&&20.73	&12.42	&0.79\%	&0.48\%\\			
& Eigen 10	&&15.13	&9.83	&0.58\%	&0.38\%\\			
& Eigen 11	&&14.58	&9.64	&0.56\%	&0.37\%\\			
& Eigen 12	&&11.67	&8.92	&0.45\%	&0.34\%\\			
& Eigen 13	&&11.27	&8.41	&0.43\%	&0.32\%\\			
& Eigen 14	&&10.60	&7.92	&0.41\%	&0.30\%\\			
& Eigen 15	&&9.71	&7.63	&0.37\%	&0.29\%\\			
							
\hline							
\hline							
\end{tabular}							
\caption{\label{DataStat} First 15 HPCA and PCA eigenvalues using statistical clustering.}					
\end{table}

\subsubsection{Eigenvector Analysis}

{}Higher-order eigenvectors are concentrated in a few clusters. The second eigenvector is not easy to associate since it is significant across the whole spectrum. See the plot below.

\begin{figure}[H]
\includegraphics[width=1\linewidth]{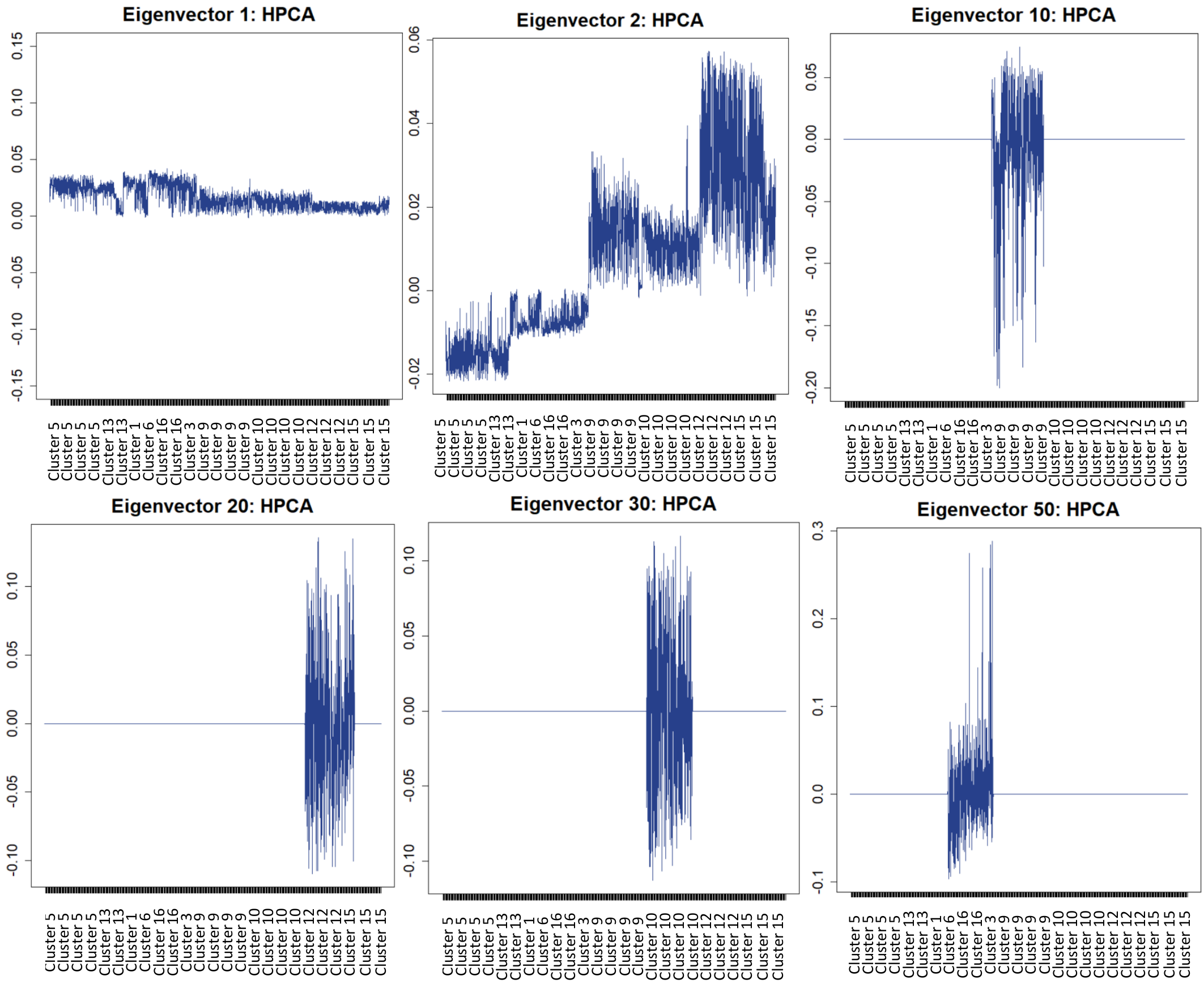}
\caption{Higher-order eigenvectors of HPCA clustered by the statistical approach. For example, the tenth eigevector is localized in \textit{Cluster 9}. The twentieth is localized mostly in \textit{Cluster 12}. The thirtieth eigenvector is totally concentrated in \textit{Cluster 10} and the fiftieth in \textit{Clusters 6 and 16}. See Table \ref{Data1Anne} for details on each cluster.}
\end{figure}

{}An interesting question at this point is oriented towards the meaning of each group.\footnote{See in the Appendix Table \ref{DataClust} the number of components of each cluster.} If the clustering technique works well, one would expect assets that share common factors, such as sectors or countries, to belong to the same cluster. However, this is not necessarily true because there are myriad ``hidden'' factors that are difficult to identify and remain important drivers of asset returns. In the end, that is the objective of the statistical clustering approach; identify clusters that are not easy to see with common factors such as sectors or countries.

{}Table \ref{Data1Anne} in the Appendix shows the main sectors and countries to which the components (assets) belong within each cluster.

{}There are two types of sectors. \textit{Cyclical sectors}: Consumer Discretionary, Financials, Real Estate, Industrials, Information Technology, Materials and Communication, and \textit{defensive sectors}:  Consumer Staples, Energy, Health Care and Utilities. 

{}The main insight obtained from this analysis is that in clusters 2, 5, 8, 9, 10, 12, 14, 15 and 16 the three main sectors are cyclical. In the other clusters, at least two out of three are also cyclical. Only cluster 13 has two defensive sectors. As for the countries, the most interesting insights are obtained from clusters 2, 4, 5, 12, 13, 14 and 15, where almost all the companies belong to China or to the US. Therefore, the statistical clustering approach is doing a good job by identifying common factors such as sector (distinguishing between cyclical and defensive sectors) and countries but also it takes into account other non-explicit (statistical) factors. This is evident in the next section, where we apply this procedure to portfolio management.

\section{Application to Portfolio Management}

{} We demonstrate how practitioners might apply HPCA to portfolio management. First we show that, after transaction costs, the main eigenportfolio has a similar performance to that of the market, as explained in Subsection \ref{FirstHP2}. Second, we use the statistical clustering approach with HPCA to build statistical factor models and build an investment portfolio.

\subsection{First Eigenportfolio \texorpdfstring{$\approx$}~~Market Portfolio}

{} We construct a portfolio based on the first eigenvector for the US stock market. The portfolio is rebalanced monthly and the correlation matrix is estimated with a rolling window of approximately 125 days.

{}As expected, the first eigenportfolio is a good proxy of the market. This confirm the points mentioned before.

\begin{figure}[H]
\centering
\includegraphics[width=0.85\linewidth]{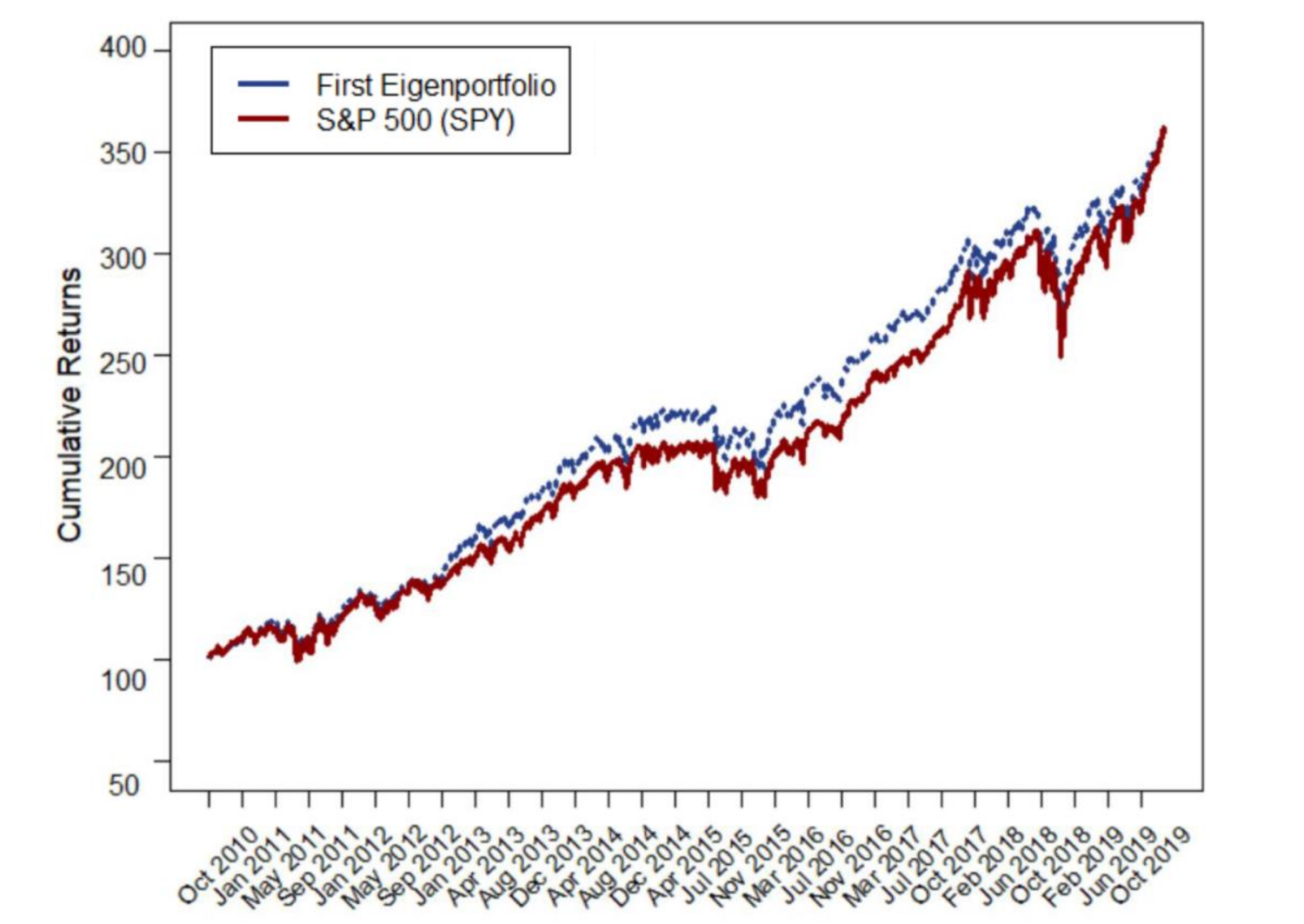}
\caption{\label{FirstEigen} Cumulative returns of the market portfolio, represented via its tradeable ETF (SPY), and the first eigenportfolio, rebalanced monthly.}
\end{figure}

\subsection{Statistical Clustering Factor Model}

{}We blend the statistical clustering approach with HPCA and use the (ordered) eigenvalues and its associated eigenvectors to build a statistical factor model for the covariance matrix and the expected returns. The model correlation matrix using $K$ components of HPCA reads
\begin{align}
& \mathcal{C}=\hat{\mathcal{O}} \hat{\Lambda} \hat{\mathcal{O}}^{T}+\zeta^{2}\\
& \zeta^2_j = \sum_{i = K + 1}^N \lambda^{(i)} (\mathcal{O}^{(i)}_j)^2 \
\end{align}
where $\zeta^2$ is the (uncorrelated) idiosyncratic risk and $\hat O$ and $\hat \Lambda$ are the modified orthogonal matrix of eigenvectors and the diagonal matrix of eigenvalues, respectively. We define the expected returns as
\begin{align}
& E(r) = \sum_{i = i}^K \beta_j^{(K)}F^{(K)} + \epsilon_i \
\end{align}
{}where $\beta_j^{(K)}$ are the factor loadings and $\epsilon_i$ the residuals.

{}To select the number $K$ of factors used to reconstruct the correlation matrix and compute the expected returns we used the so-called effective rank (eRank) method. Let the singular value decomposition of the $T \times N$ matrix of standardized log-return
\begin{align}
& R = UDV \
\end{align}
{}where $U$ and $V$ are $T\times T$ and $N \times N$ unitary matrices and $D$ is the diagonal matrix with singular values in decreasing order. Then, the associated probability distribution is 
\begin{align}
& \mathcal{P}_j = \frac{\sigma_j}{\Vert \sigma \Vert_1} ~~~ \text{for} ~~~j = 1,...,Q ~~~ \text{for}~~~ Q = M \wedge T\
\end{align}
{}where $\Vert . \Vert_1$ is the $L-1$ norm. Then, the effective rank is defined as 
\begin{align}
& eRank(R) = exp\{\mathcal{H}(\mathcal{P}_1, \mathcal{P}_2,...,\mathcal{P}_Q)\} \
\end{align}
{}where $\mathcal{H}$ is the Shannon entropy.\footnote{See \cite{KakushadzeYu2017} for more implementation details.}

\begin{figure}[H]
\centering
\includegraphics[width=0.85\linewidth]{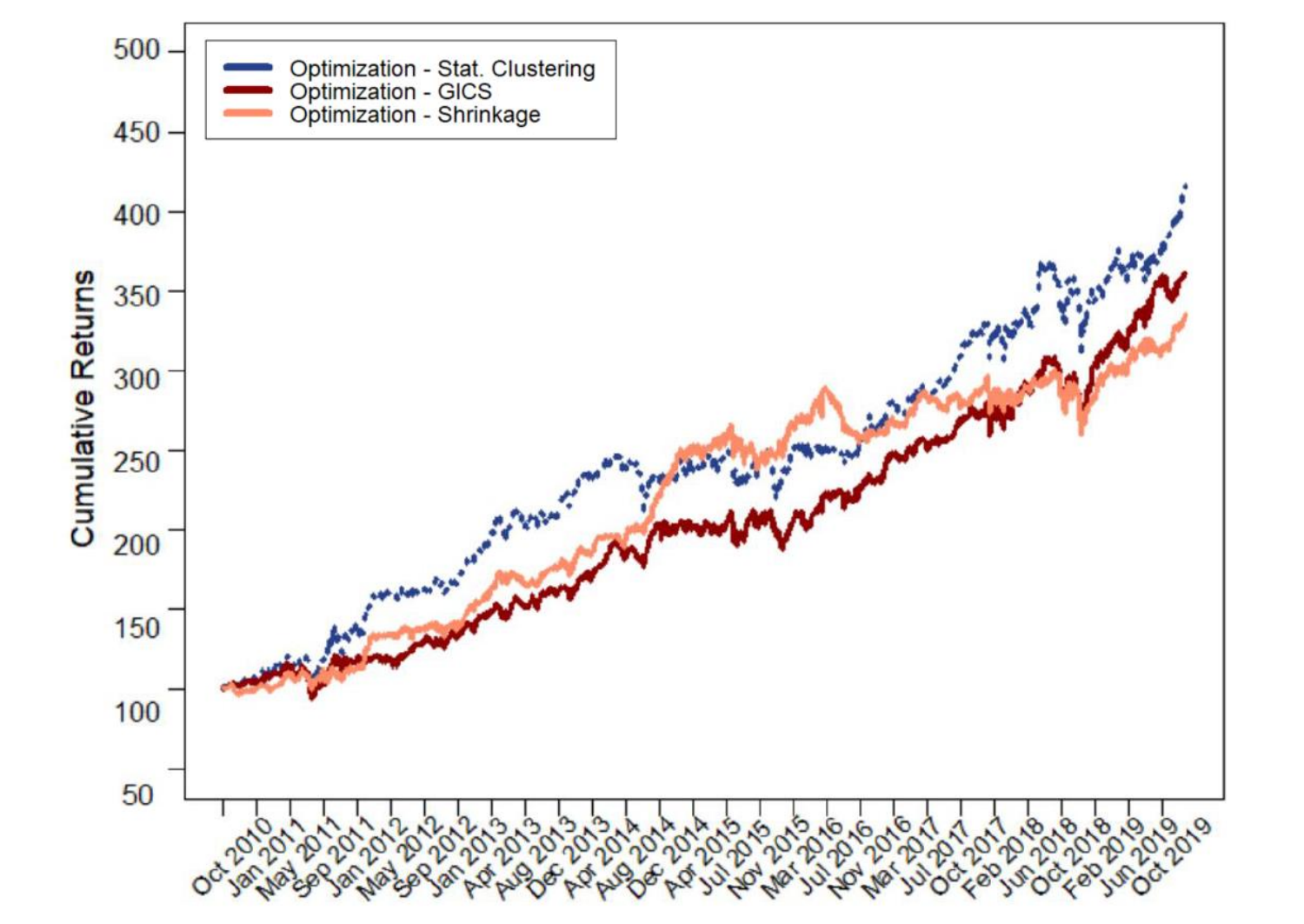}
\caption{\label{Optimization} In this plot we show the cumulative returns of three optimized portfolios based on HPCA. The blue line is the strategy performed using the statistical clustering approach, the red line represents the strategy based on the GICS clustering and salmon line represents the classical Sharpe ratio optimization with shrinkage covariance matrix.}
\end{figure}

{}Figure \ref{Optimization} shows the cumulative returns of the different strategies analyzed. 

{}Table \ref{strats} displays the main performance statistics of the three optimization techniques, the passive S\&P 500 Index and the First Eigenportfolio, showing that the optimization approach based on statistical clustering with HPCA outperforms all the other portfolios.

\begin{table}[H]
\centering
\caption{\label{strats}Main performance statistics of the proposed strategies. The risk-free rate was set to zero.}
\begin{tabular}{lccccc}
\hline

 & CAGR & Std Dev & Sharpe ratio & MaxDD & Calmar ratio \\ \hline
First Eigen & 15.36\% & 14.31\% & 1.07 & 21.7\% & 0.71 \\ 
Stat. Clustering & 16.91\% & 13.65\% & 1.24 & 16.3\% & 1.04 \\ 
GICS & 15.13\% & 13.06\% & 1.16 & 18.0\% & 0.84 \\ 
Shrinkage & 14.19\% & 11.48\% & 1.24 & 14.7\% & 0.97 \\ 
S\&P 500 & 15.13\% & 13.99\% & 1.08 & 19.9\% & 0.76 \\ \hline \hline

\end{tabular}
\end{table}

\newpage
\section*{Conclusions}

{}Throughout this paper, we have empirically compared the performance of PCA and HPCA. We tested different approaches for more than 2600 stocks, grouping them by market regions: United States, Europe, China, and Emerging stock markets. Main results provide evidence that HPCA works remarkably well for different markets, tackling one of the main shortcomings of classical PCA, the so-called identification problem. The HPCA portfolios, unlike the PCA ones, are easy to interpret and embeds an economic/financial intuition behind. Thus, these portfolios can be used as factors to build models for risk and portfolio management, as well as for different trading strategies such as statistical arbitrage.

{}Clustering stock returns by GICS and countries provides outstanding results when HPCA is applied. However, these types of clustering approaches have some drawbacks. First, the user needs to specify the clusters, which are not always available for all markets. Second, static groups like these cannot quickly adjust to changes in market conditions. In this matter, we showed that under different market regimes, the correlation between different assets changes sharply and so, the behavior of returns. This, in turn, could affects the covariance matrix estimators and the performance of well-known trading strategies such as factor/cluster rotation. A seemingly well-diversified portfolio can get concentrated in a few components if market conditions change and the user does not account for that timely.

{}To tackle these issues inherent in static clusters, we propose a purely statistical cluster approach, in which, the user no longer needs to specify the clusters, since they are obtained from the spectral decomposition of the correlation matrix of the stock returns. This approach is more promising to account for changes in the behavior of assets. Furthermore, by definition this tool is parsimonious provided that the user only needs to fix up one parameter, the number of $K$ eigenvectors.\footnote{Although not covered in this paper, the user could use some heuristics to set the $K$ number of eigenvectors.} To illustrate an application, we show it in the context of portfolio optimization for the US stock market. We provide evidence that using HPCA statistical-based factor models outperform other classical portfolio construction methodologies such as the shrinkage covariance matrix and the HPCA GICS-based factor models.

\pagebreak
\section*{Appendix}
\begin{table}[H]
\centering
\begin{tabular}{rlrlr}
  \hline
 & Europe & Number & Emerging & Number \\ 
  \hline
& United Kingdom & 147 & China & 288 \\ 
& France &  89 & Korea & 114 \\ 
& Germany &  77 & Taiwan &  87 \\ 
& Switzerland &  52 & India &  80 \\ 
& Sweden &  44 & Brazil &  55 \\ 
& Italy &  33 & Hong Kong &  50 \\ 
& Spain &  26 & South Africa &  44 \\ 
& Netherlands &  25 & Malasya &  43 \\ 
& Denmark &  21 & Thailand &  37 \\ 
& Belgium &  18 & Saudi Arabia &  31 \\ 
   \hline
&\textbf{Total} &  \textbf{532} & & \textbf{829}\\
\hline
\hline
\end{tabular}
\caption{\label{DataCountries} Number of companies by countries in European and Emerging markets.}
\end{table}

\begin{table}[H]
\centering
\begin{tabular}[t]{clcccc}
   \hline
  \hline
& Cluster && Number && Percentage (\%) \\ 
  \hline
&Cluster 1	&&24     &&0.91\%\\
&Cluster 2	&&9	     &&0.34\%\\
&Cluster 3	&&90	 &&3.41\%\\
&Cluster 4	&&18	 &&0.68\%\\
&Cluster 5	&&368	 &&13.94\%\\
&Cluster 6	&&194	 &&7.35\%\\
&Cluster 7	&&1	     &&0.04\%\\
&Cluster 8	&&20	 &&0.76\%\\
&Cluster 9	&&406	 &&15.38\%\\
&Cluster 10	&&352	 &&13.33\%\\
&Cluster 11	&&1	     &&0.04\%\\
&Cluster 12	&&388	 &&14.70\%\\
&Cluster 13	&&207	 &&7.84\%\\
&Cluster 14	&&82	 &&3.11\%\\
&Cluster 15	&&127	 &&4.81\%\\
&Cluster 16	&&353	 &&13.37\%\\
\hline
\hline
\end{tabular}
\caption{\label{DataClust} Number of assets that belong to each cluster. Some clusters, such as \textit{Cluster 7} and \textit{Cluster 11}, only have one component.}
\end{table}

\newpage

\pagebreak
\begin{longtable}{ | l | M | c | M | c |}
    \hline
    Cluster & Main Sectors & Sectors (\%) & Main Countries & Countries (\%)\\ \hline
    Cluster 1 & Materials; Health Care; Consumer Discretionary   & 50.00\%  & United Kingdom; Germany; Italy & 62.5\% \\ \hline
    & \\[\dimexpr-\normalbaselineskip+3pt]
    Cluster 2 & Materials;  Financials; Industrials & 77.78\% & China; Germany; Italy  & 77.78\% \\ \hline
    & \\[\dimexpr-\normalbaselineskip+3pt]
    Cluster 3 & Materials;  Financials; Consumer Staples & 48.31\% & China; India; Korea  & 34.83\% \\ \hline
      \\[\dimexpr-\normalbaselineskip+2pt]
    Cluster 4 & Financials; Health Care; Consumer Discretionary & 61.11\% & China; Brazil; Korea  & 72.22\% \\ \hline
    & \\[\dimexpr-\normalbaselineskip+3pt]
    Cluster 5 & Information Technology;  Industrials; Consumer Discretionary & 50.81\% & United States; United Kingdom; Ireland & 97.55\% \\ \hline
    & \\[\dimexpr-\normalbaselineskip+3pt]
    Cluster 6 & Industrials;  Financials; Consumer Staples & 42.48\% & United Kingdom; Germany; France  & 37.15\% \\ \hline
    & \\[\dimexpr-\normalbaselineskip+3pt]
    Cluster 8 & Materials;  Financials; Information Technology & 55.00\% & United Kingdom; Germany; France  & 50.00\% \\ \hline
    & \\[\dimexpr-\normalbaselineskip+3pt]
    Cluster 9 & Materials;  Financials; Consumer Discretionary & 49.62\% & China; Taiwan; Korea  & 39.00\% \\ \hline
    & \\[\dimexpr-\normalbaselineskip+3pt]
    Cluster 10 & Materials;  Financials; Information Technology & 47.57\% & China; India; Korea  & 43.33\% \\ \hline
    & \\[\dimexpr-\normalbaselineskip+3pt] 
    Cluster 12 & Materials;  Industrials; Information Technology & 53.35\% & China; Taiwan; Hong Kong  & 96.91\% \\ \hline    
    & \\[\dimexpr-\normalbaselineskip+3pt]
    Cluster 13 & Consumer Staples;  Real Estate; Utilities & 49.10\% & United States; Brazil; China  & 80.01\% \\ \hline
    & \\[\dimexpr-\normalbaselineskip+3pt]
    Cluster 14 & Consumer Discretionary; Communication; Financials & 49.41\% & China; Korea; Thailand  & 75.60\% \\ \hline
    & \\[\dimexpr-\normalbaselineskip+3pt]
    Cluster 15 & Consumer Discretionary;  Information Technology; Industrials & 49.20\% & China; Indonesia; Malaysia & 96.03\% \\ \hline
    & \\[\dimexpr-\normalbaselineskip+3pt]
    Cluster 16 & Financials; Industrials; Consumer Discretionary  & 55.24\% & United Kingdom; France; Germany  & 52.11\% \\ \hline
    \hline
    \caption{\label{Data1Anne} Main countries and industries belonging to each cluster.}
\end{longtable}

\end{document}